\documentclass[namedreferences]{solarphysics}

\usepackage[hyperref,optionalrh,showbiblabels]{spr-sola-addons} 
\usepackage{graphicx}        
\usepackage{color}           

\newcommand{\BibTeX}{\textsc{Bib}\TeX}

\renewcommand{\vec}[1]{{\mathbfit #1}}
\newcommand{\deriv}[2]{\frac{{\mathrm d} #1}{{\mathrm d} #2}}
\newcommand{\rmd}{ {\ \mathrm d} }

\newcommand{\vol}{ {\mathcal V} }

\newcommand{\dv}{~{\mathrm d}^3 x}

\newcommand{\intv}{\int_{\vol}^{}}

\newcommand{\avec}{ \vec A}

\newcommand{\jj}{ \vec j}

\newcommand{\xx}{ \vec x}

\newcommand{\BB}{\vec B} 
\newcommand{\bb}{\vec b} 
\newcommand{\bv}{\vec v} 
\newcommand{\np}{n_{p}} 
\newcommand{\elsasser}{\vec z^{\pm}} 
\newcommand{\elsasserplus}{z^{+}} 
\newcommand{\elsasserminus}{z^{-}} 
\newcommand{\vsw}{V_{sw}} 
\newcommand{\sigmac}{\sigma_{c}}
\newcommand{\sigmar}{\sigma_{r}}
\newcommand{\deltat}{\Delta t}
\newcommand{\aap}{    {\it Astron. Astrophys.}}

\newcommand{\apj}{    {\it Astrophys. J.}}

\newcommand{\jgr}{    {\it J. Geophys. Res.}}

\newcommand{\mnras}{  {\it Mon. Not. Roy. Astron. Soc.}}

\newcommand{\pasp}{   {\it Pub. Astron. Soc. Pac.}}
\newcommand{\pasj}{   {\it Pub. Astron. Soc. Japan}}

\newcommand{\solphys}{{\it Solar Phys.}}

\chardef\us=`\_

\begin{document}

\begin{article}

\begin{opening}

\title{Turbulence, intermittency and cross-scale energy transfer in an interplanetary coronal mass ejection}

\author{\inits{R.}\fnm{Roque}~\lnm{M\'arquez Rodr\'iguez}$^{1,2,3}$} \author[corref,email={lucasorriso@gmail.com}]{\inits{L.}\fnm{Luca}~\lnm{Sorriso-Valvo}$^{4,3}$\orcid{0000-0002-5981-7758}} \author{\inits{E.}\fnm{Emiliya}~\lnm{Yordanova}$^{3}$\orcid{0000-0002-9707-3147}}




\institute{$^{1}$ Facultad de F\'isica, Universidad de Santiago de Compostela (USC), Calle Xos\'e Mar\'ia Su\'arez N\'u\~nez, s/n, 15782 Santiago de Compostela, Spain}
\institute{$^{2}$ Observational Astrophysics, Division of Astronomy and Space Physics, Department of Physics and Astronomy, Uppsala University, Box 516, 75120 Uppsala, Sweden}
\institute{$^{3}$ Swedish Institute of Space Physics (IRF), \r{A}ngstr\"om Laboratory, L\"agerhyddsv\"agen 1, SE-751 21 Uppsala, Sweden}
\institute{$^{4}$ CNR/ISTP -- Istituto per la Scienza e Tecnologia dei Plasmi, Via Amendola 122/D, 70126 Bari, Italy}


\runningauthor{M\'arquez et al.}
\runningtitle{Turbulence in a ICME}

\begin{abstract}

Solar wind measurements carried out by NASA's Wind spacecraft before, during and after the passing of an interplanetary coronal mass ejection (ICME) detected on 12-14 September 2014 have been used in order to examine several properties of magnetohydrodynamic (MHD) turbulence. 
Spectral indices and flatness scaling exponents of magnetic field, velocity and proton density measurements were obtained, and provided a standard description of the characteristics of turbulence within different sub-regions of the ICME and its surroundings. 
This analysis was followed by the validation of the third-order moment scaling law for isotropic, incompressible MHD turbulence in the same sub-regions, which confirmed the fully developed nature of turbulence in the ICME plasma. 
The energy transfer rate was also estimated in each ICME sub-region and in the surrounding solar wind. An exceptionally high value was found within the ICME sheath, accompanied by enhanced intermittency, possibly related to the powerful energy injection associated with the arrival of the ICME.
\end{abstract}


\end{opening}

\section{Introduction}
     \label{S-Introduction}

The dynamics of the solar wind, a plasma flow expanding from the Sun through the whole heliosphere at supersonic and super-Alfv\'enic speed \citep{Parker1958}, has been extensively explored by spacecraft measurements. 
Its temporal and radial evolution remains hard to predict, due to the non-linear interactions and turbulent behaviour that characterize its fluctuations. 
The turbulent nature of the solar wind is a major subject of space plasma physics research, and has been studied in depth for more than 50 years \citep[][]{Viall2020}. 
Power-law spectra and anomalous scaling of the structure functions of magnetic field, velocity and proton density have been widely used to characterize the turbulence and the associated intermittency \citep[for an exaustive account of solar wind turbulence, see the excellent review by][and references therein]{BrunoCarbone2013}. 
However, due to the complexity and variability of the solar wind environment, several aspects of solar wind turbulence are still being investigated. 
Understanding the evolution and properties of solar wind turbulence is of paramount importance to determine how the solar wind collisionless plasma is heated during its expansion, and for the transport of energy, momentum and other invariants in the heliosphere \citep[][]{Matthaeus2011}. 

The complexity of solar wind dynamics is further exacerbated in the case of violent transient phenomena, such as interplanetary coronal mass ejections \citep[hereafter ICMEs,][]{Howard2011,Kilpua2017}. 
These are powerful events of solar origin consisting of the expulsion of plasma and magnetic field from the corona (coronal mass ejections, CMEs), which then expand with high speed through the interplanetary space. 
Their high speed, often supersonic with respect to the embedding solar wind, produces a shock wrapped around the expanding magnetic ejecta, which causes compression and heating of the ambient plasma. 
The normal structure of ICMEs consists of a highly compressed and turbulent sheath immediately behind the shock, followed by a colder, quieter magnetic cloud, which represents the bulk of the expelled plasma. 
The speed, size, geometry and magnetic configuration of ICMEs can be extremely variable, and so do their dynamical properties. 
Furthermore, the interaction with the inhomogeneous solar wind or other ICMEs also contributes to determine their expansion speed and other characteristics \citep{Dallago2003,Wang2004,Gui2011,Lugaz2017,Heinemann2019,Wang2020}. 
They represent therefore an exceptionally complex system for theoretical modeling and experimental studies. 
Understanding the way ICMEs propagate from the Sun in the interplanetary space is also a crucial ingredient of space weather \citep{Temmer2021}. Indeed, when ICMEs reach the near-Earth space and the terrestrial magnetosphere, the perturbations they produce in the solar wind-magnetosphere coupling can originate harmful space weather events \citep{Schwenn2005,Bothmer2007,Pulkkinen2007,Echer2013,Kilpua2017}. 
The possibility of accurate modeling for the prediction of time of arrival and conditions of ICMEs impacting the Earth heavily relies on the knowledge of their complex dynamics.  

For example, recent studies have highlighted that the interplanetary plasma turbulence is severely affected by the interaction with ICMEs \citep[see, e.g.,][]{Kilpua2021}, which in turn feeds back on the ICME propagation. 
However, the interplay of ICMEs with the ambient wind turbulence is still largely unexplored \citep{Sorriso-Valvo2021}.
The aim of this paper is to analyze the turbulent properties of ICMEs and of the preceding and trailing solar wind, using one case study measured on 12-14 September 2014 by the Wind spacecraft, and not yet presented in the literature. 
The analysis will be based on the scaling of the structure functions and on the Politano-Pouquet law for isotropic, incompressible magnetohydrodynamic (MHD, hereafter) plasmas \citep{Politano1998}. 
This work represents in fact an extension of the analysis performed by \citet{Sorriso-Valvo2021} using a different ICME measured in 2012, and a contribution towards a more complete characterization of turbulence within and around ICMEs.

The structure of this paper is as follows: Section \ref{S-Description of the data} presents the experimental data and the selected sub-intervals for the study of turbulence; Section \ref{S-Structure function-based analysis} provides the results of the two-points structure function analysis; Section \ref{S-Third-order moment scaling law} contains the analysis of the Politano-Pouquet law and of the mean energy transfer rate; finally, Section \ref{S-Conclusion} ends with a summary and discussion of the results.

\section{Description of the data}
     \label{S-Description of the data}

A fast interplanetary coronal mass ejection was measured between 12 and 14 September 2014 by NASA's Wind spacecraft. Its turbulent properties and those of the preceding and trailing solar wind will be studied through plasma moment and magnetic field measurements. Proton velocities and densities are indicated by $\bv$ and $\np$, respectively, while $\BB$ represents the magnetic field. The magnetic field measurements were carried out by Wind's Magnetic Field Investigation (MFI) magnetometer \citep{Lepping1995}. 
All vector quantities are expressed in the GSE coordinate system (i.e., pointing the $z$-axis to the ecliptic north and being the $x$-axis directed from the Earth to the Sun). 

\begin{figure}[h!]
\centering
\includegraphics[width=0.98\textwidth]{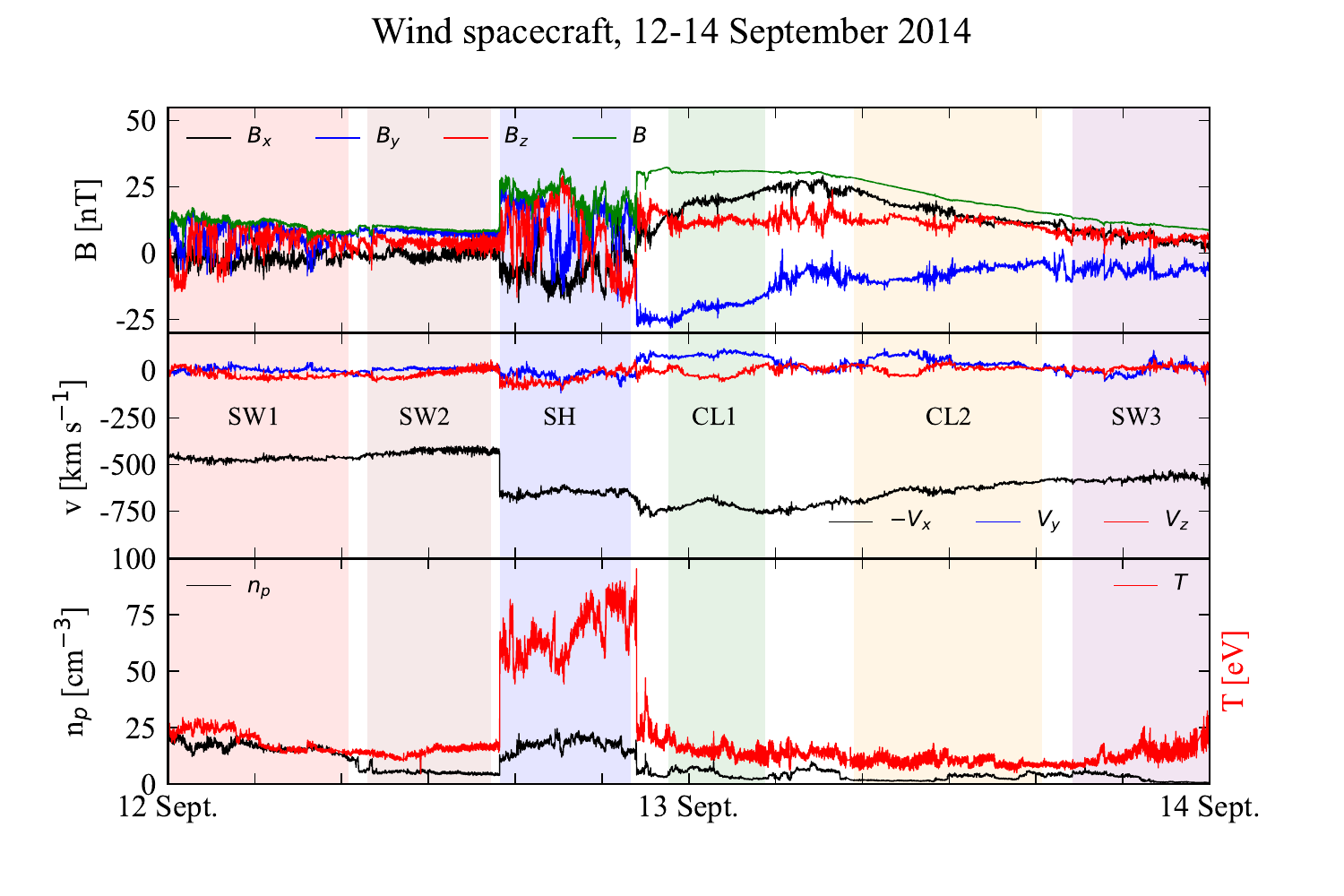}

\caption{ Magnetic field components and magnitude (top panel), proton velocity components (middle panel) and proton densities and temperatures (bottom panel; the $y$-axis label scale is the same for both $n_{p}$ and $T$). Coloured areas represent six selected intervals in terms of their relative homogeneity; their labels are found in the middle panel and the colour code employed here will be used in the subsequent graphs.}
\label{windmeasurements}
\end{figure}

Measured solar wind parameters during the event are shown in Figure \ref{windmeasurements}. 
The top panel displays values of magnetic field components and magnitude, downsampled to the plasma cadence. 
The middle panel shows measurements of proton velocities. 
The third panel shows proton density and temperature measured values\footnote{Note that a few instances of artificial spikes were removed manually from the data. In addition, the public database contains a long sequence of repeated timestamps, associated with constant values of the plasma moments, which have been also manually removed from our sample.}. 
From the figure, it emerges that the structure of the measured ICME is very clear, and the different regions are well separated and recognisable.
The interplanetary shock associated with the ICME arrives at WIND spacecraft at 15:28 UTC on 12 September and on Figure \ref{windmeasurements} it is clearly visible by the sharp gradients in the magnetic field and plasma parameters. 
Behind the shock, between 15:28/2014-09-12 and 21:33/2014-09-12 UTC, follows the ICME sheath consisting of compressed and heated solar wind that has been piled-up from the expanding magnetic ejecta. It is seen as the high density and temperature interval relative to the adjacent regions. Behind the sheath arrives the magnetic cloud of the ICME (23:05/2014-09-12 - 16:27/2014-09-13 UTC), which has strong magnitude but very low level of fluctuating and slowly rotating magnetic field. 
The magnetic cloud's plasma is very cool and low density, which is typically the in-situ signature of the magnetic ejecta \citep{Zurbuchen2006}. 

Based on such structure, Figure \ref{windmeasurements} displays six different colour-shaded sub-intervals, which identify regions in the event with relatively homogeneous statistical properties. 
Such regions include: 
(1) a pristine solar wind sample (SW1, red) of average bulk velocity $\vsw = 471$ km s$^{-1}$, followed by 
(2) another quieter solar wind interval in the proximity of the ICME shock (SW2, brown, with $\vsw= 431$ km s$^{-1}$); 
(3) a region of intense fluctuations downstream of the evident shock, corresponding to the ICME sheath (SH, blue, $\vsw = 652$ km s$^{-1}$); 
(4-5) an ICME cloud sample divided into two subintervals, one being closer to the leading cloud's edge (CL1, green, $\vsw = 724$ km s$^{-1}$) and the other corresponding to the trailing part of the cloud (CL2, yellow, $\vsw = 635$ km s$^{-1}$); CL1 is characterized by an almost constant magnetic field magnitude and CL2 by its smooth decay and that of the velocity, indicating the ICME expansion; note that the two cloud sub-intervals are separated by a broad region of relative more intense fluctuations, which was excluded from the analysis; 
(6) a solar wind sample beyond the ICME (SW3, purple), with mean velocity $\vsw = 577$ km s$^{-1}$. 
Care has been taken to ensure that all of them are long enough in order to be statistically accurate, via a standard autocorrelation function-based analysis and a convergence study \citep{Dudok2013}. 
%

\section{Structure function-based analysis of turbulence and intermittency}
     \label{S-Structure function-based analysis}

In magnetized plasmas such as the solar wind, the interplay between inertial-range turbulence and large-scale Alfv\'enic fluctuations requires the introduction of parameters that quantify both properties.

Alfv\'enic fluctuations indicate highly aligned (either correlated or anticorrelated) velocity and magnetic field fluctuations, typically associated with Alfv\'en waves propagating along the ambient magnetic field. 
These are conveniently studied using the Elsasser variables, $\elsasser = \bv \pm \vec \bb$ \citep{Elsasser1950}, where $\bv$ and $\bb$ stand for velocity and magnetic field (transformed in velocity units through $\bb =\BB / \sqrt{4 \pi \np m_{p}}$, being $m_{p}$ the proton mass). 
The Alfv\'enic nature of the fluctuations can be assessed using the cross-helicity, or the mean $\bv$-$\bb$ alignment, an invariant of the incompressible MHD equations defined as $H_c=\langle \bv \cdot \bb \rangle$, where brackets indicate ensemble average.
Dividing by the incompressible fluctuation energy density per unit mass, $E=\left\langle|\bv|^2+|\bb|^2\right\rangle / 2$, gives the normalized cross-helicity, $\sigma_c=H_c / E$, whose values lie between -1 and 1 \citep[see for example][and references therein]{BrunoCarbone2013}.
The balance between magnetic and velocity fluctuations is also described using the residual energy, $\sigmar$, which can be expressed as $\sigma_r=\left(\left\langle|\bv|^2\right\rangle-\left\langle|\bb|^2\right\rangle\right) /\left(\left\langle|\bv|^2\right\rangle+\left\langle|\bb|^2\right\rangle\right)$ \citep{BrunoCarbone2013}.


Standard turbulence models are broadly based on the Kolmogorov phe\-no\-me\-no\-lo\-gi\-cal description (K41) of the turbulent cross-scale energy transfer due to the nonlinear interactions among fields fluctuations \citep{Kolmogorov1941}. 
If the turbulence is fully developed, the nonlinear energy transfer (stemming from the nonlinear term of the fluid dynamical equations) is the dominant process in the so-called inertial range of scales, where both the energy large-scale injection and small-scale dissipation can be neglected. Such transfer generates a cascade of energy from large to small scale, where it is eventually dissipated \citep{Frisch1995}.
In the inertial range, the incompressible magnetohydrodynamics (MHD) equations are invariant under scaling transformations, so that the fields fluctuations have power-law scaling, $\Delta \phi \sim \ell^h$, where $\Delta \phi=\phi(t+\ell)-\phi (t)$ represent two-point increments of a scalar or field component, $\phi$, across a scale $\ell$ and $h$ is the scaling exponent that determines the statistical properties of the fluctuations. 
Based on dimensional arguments, the K41 phenomenology predicts the scaling exponent $h=1/3$.
The equations' scale invariance results in kinetic and magnetic spectra that decay as a power-law of the wavevector, $E(1/\ell) \sim (1/\ell)^{-\alpha}$, with the K41 spectral exponent $\alpha=2h+1=5/3$. 
In magnetized plasmas, the presence of large-scale Alfv\'en waves may slow down the nonlinear interactions and reduce the turbulence. 
In this case, phenomenology provides a shallower scaling exponent, $h=1/4$, and the corresponding Iroshnikov-Kraichnan (IK) spectral exponent, $\alpha=3/2$ \citep{Iroshnikov1964,Kraichnan1965}. 
An important feature universally observed in turbulent flows is intermittency. 
Since the scaling exponent $h$ is not necessarily constant, the energy transfer across scale is spatially inhomogeneous, resulting in the progressive concentration of energy on small-scale structures that are intermittently distributed in the volume \citep{Kolmogorov1962}. 
Such inhomogeneity is associated with the scale-dependent statistical properties of the fluctuations, whose distribution function changes from Gaussian at large scale to high-tailed at small scale, accounting for the small-scale energy accumulation in strong structures.  
The scale-dependent $q$-order moments of the fluctuations, called structure functions, $S_{q}(\Delta t)=\left\langle \Delta \phi^{q} \right\rangle$ (where $\Delta \phi=\phi(t+\Delta t)-\phi (t)$, represent two-point increments across a timescale $\Delta t$) provide a basic tool to study the scale-dependent statistical properties of turbulent fluctuations \citep{Frisch1995}. 
Customarily, for a time series of turbulent flows the Taylor hypothesis \citep{Taylor1938} links timescales, $\deltat$, with length scales, $\ell$, via the simple relation $\ell=-\vsw \deltat$, so that time increments can be used to describe the turbulent statistical properties. The Taylor hypothesis is robustly valid in all samples under study.
From the scaling properties of the MHD variables (e.g., velocity and magnetic field), in the inertial range the structure functions have power-law scaling, $S_{q}(\Delta t) \sim \Delta t^{\zeta_q}$. 
In the K41 or IK description, for which $h$ is constant, the scaling exponents increase linearly with the structure function order, $\zeta_q\sim hq$. Deviation from such linear relation, referred to as anomalous scaling of the structure functions, indicates intermittency.
The anomalous scaling exponents, $\zeta_q$, are commonly used to quantitatively characterize the intermittency \citep{Frisch1978,Frisch1995}.
Both K41 and IK models, with their intermittent corrections, provide basic descriptions of the statistical properties of turbulent fluctuations. More complete specific descriptions for the solar wind, not used in this work, include the effects of the anisotropy imposed by the large-scale magnetic field and by the radial expansion \citep[e.g., see][and references therein]{Oughton2020}.

In this study, the scale-dependent structure functions will be used in order to obtain information on the Alfv\'enic properties of the system, the turbulent energy spectra, and the intermittent character of the fluctuations.

We start by analyzing the Alfv\'enic properties of the fluctuations, by means of the structure function-based normalized cross-helicity, $$\sigmac = [S_{2}(\elsasserplus)-S_{2}(\elsasserminus)]/[S_{2}(\elsasserplus) + S_{2}(\elsasserminus)]\, ,$$ and residual energy, $$\sigmar = [S_{2}(v)-S_{2}(b)]/[S_{2}(v)+S_{2}(b)]\, ,$$ where arguments stand for the traces of the corresponding vectors, e.g. $b = (1/3)(b_{x}+b_{y}+b_{z})$ \citep[][]{BrunoCarbone2013}. 

\begin{figure}    
   \centering{
               \includegraphics[width=0.49\textwidth,clip=]{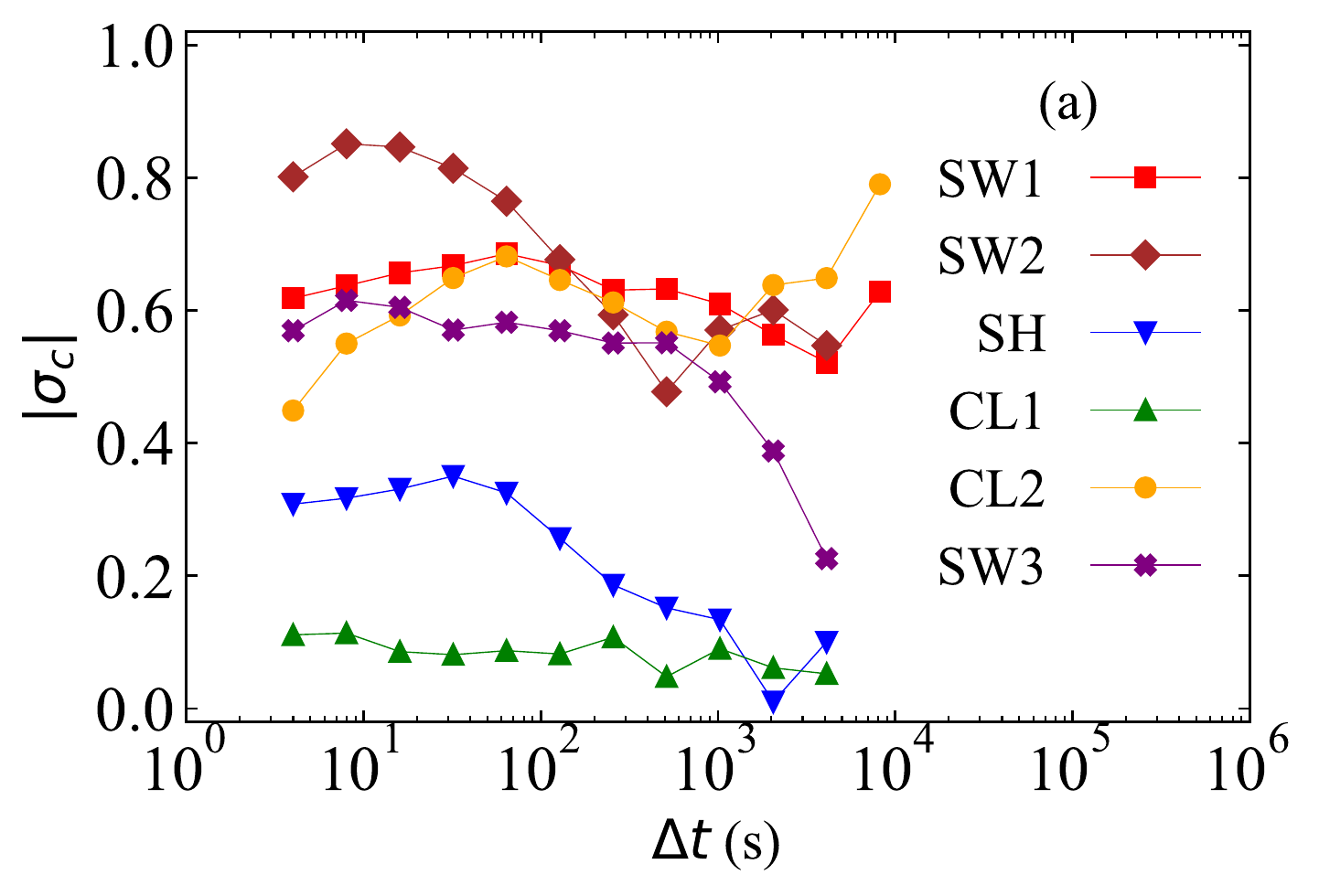}\includegraphics[width=0.49\textwidth,clip=]{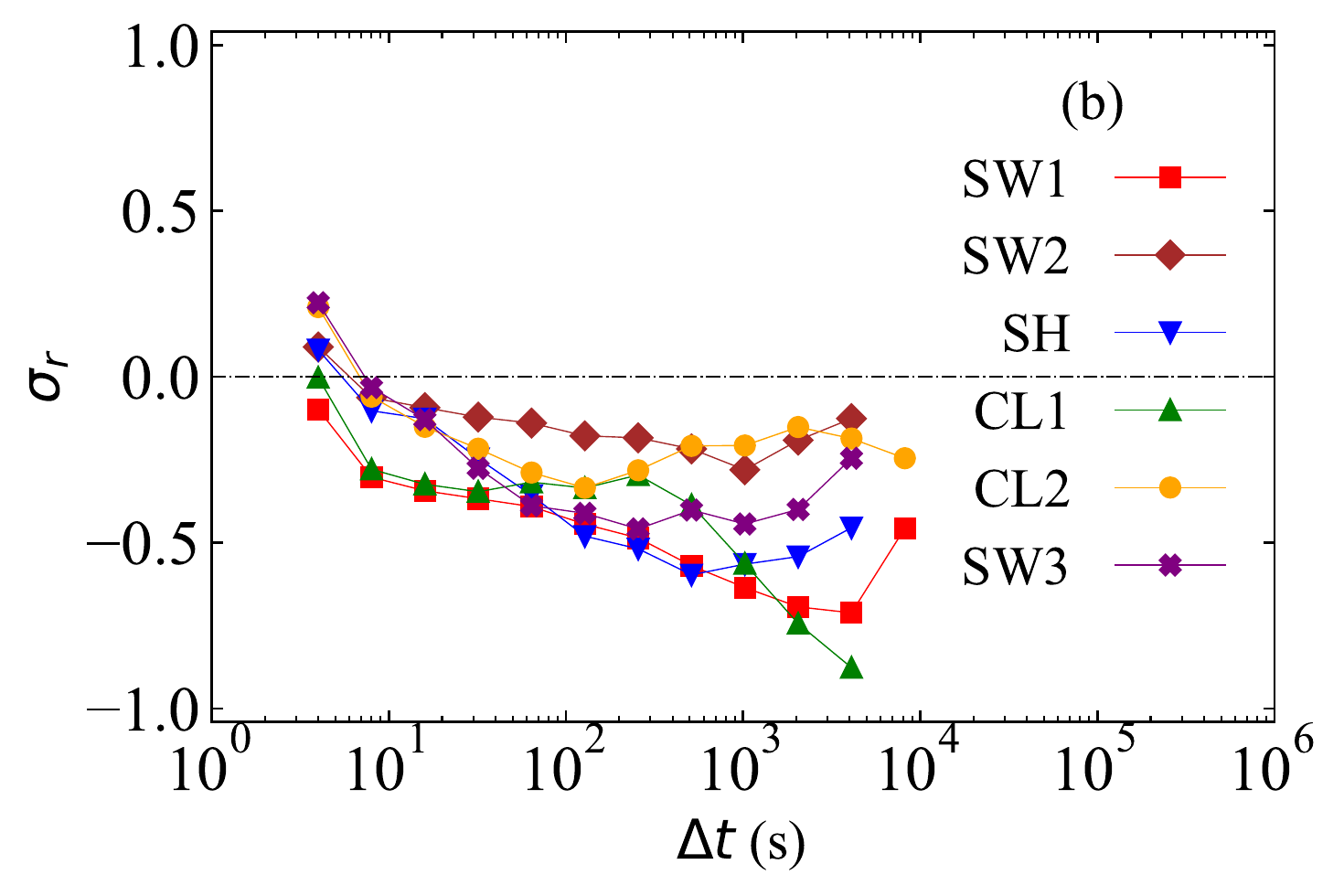} 
              }

\caption{Structure function-based normalized cross helicity $\sigmac$ (panel (a); absolute values) and residual energy $\sigmar$ (panel (b)) plotted against different timescales $\deltat$ in each region.}
   \label{F-sigmafigures}
   \end{figure}

Normalized cross-helicity and residual energy values have been plotted in Figure \ref{F-sigmafigures} against several timescales $\deltat$ for each interval. 
From the cross-helicity scaling (panel (a)), it appears that the solar wind before and after the ICME has relatively strong Alfv\'enic correlations. On the other hand, the sheath and the first cloud segment have limited correlations, as expected for the highly compressed plasma downstream of the shock. 
The negative residual energy (panel (b)) indicates that the turbulence is characterized by strong magnetic fluctuations. Both results are in agreement with those of \citet{Sorriso-Valvo2021}, suggesting that the overall Alfv\'enic nature of the turbulent fluctuations of the two ICME is similar.

\begin{figure}    
    \centerline{\includegraphics[width=0.49\textwidth,clip=]{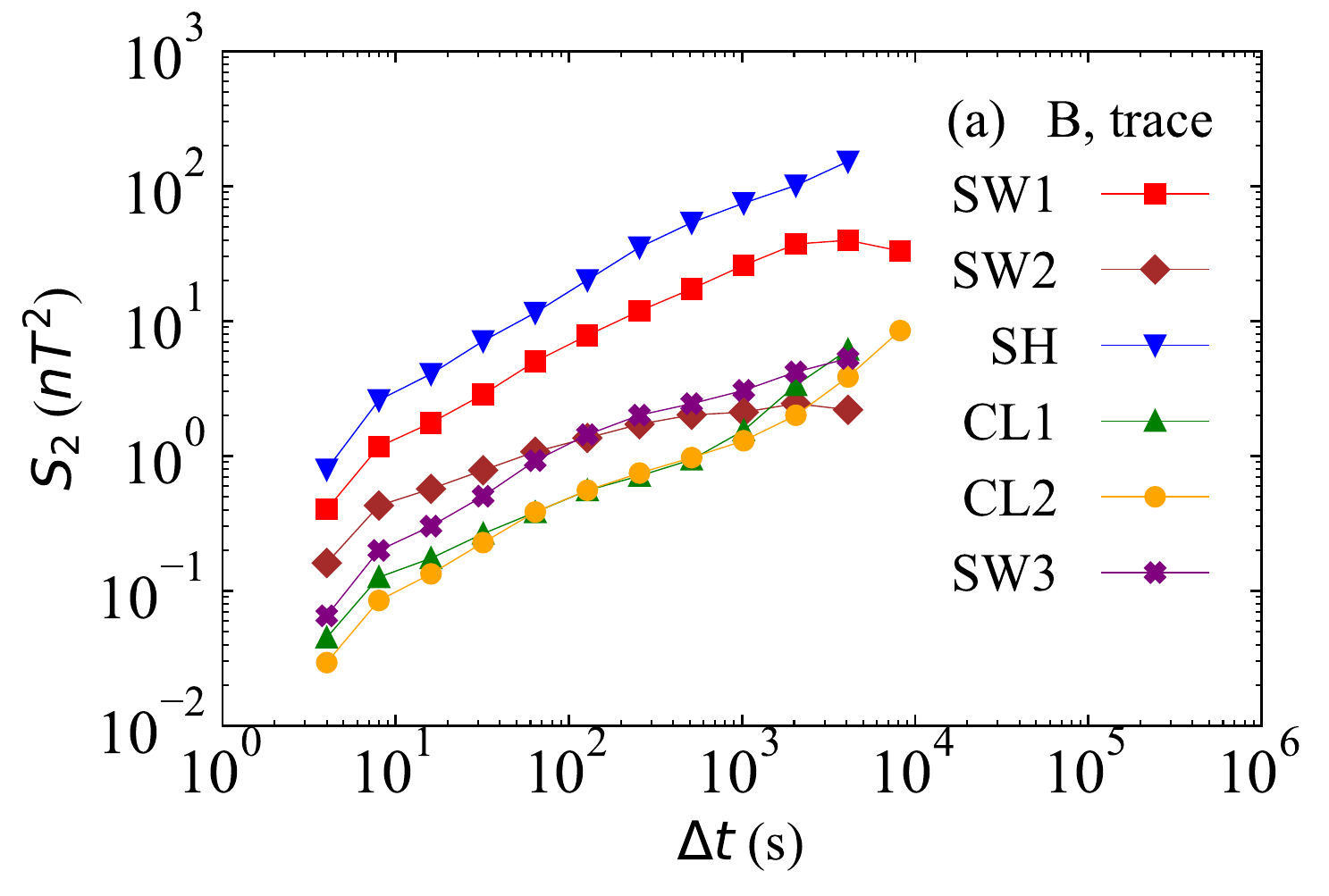} \includegraphics[width=0.48\textwidth,clip=]{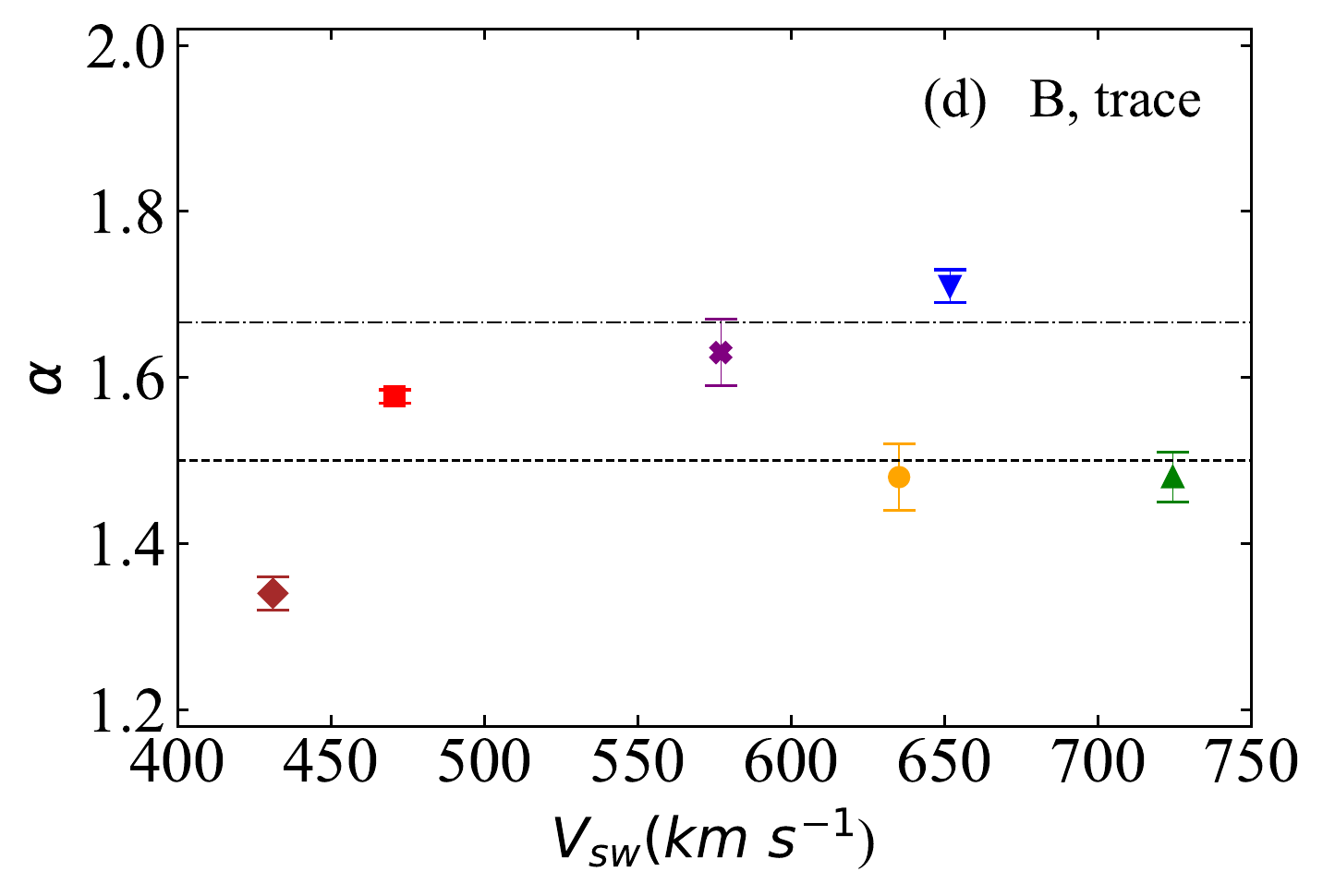}
    }
    \centerline{\includegraphics[width=0.49\textwidth,clip=]{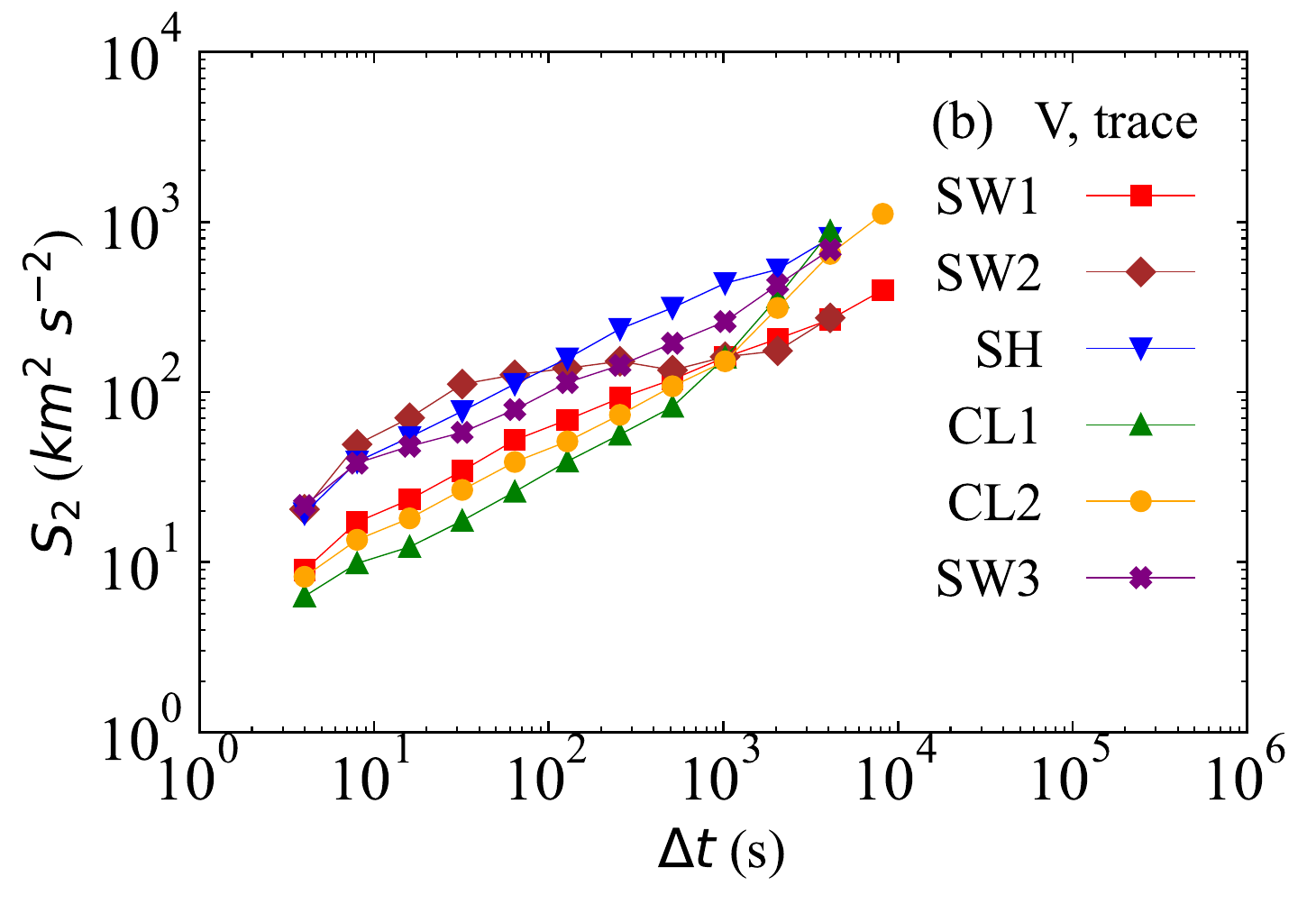}
    \includegraphics[width=0.49\textwidth,clip=]{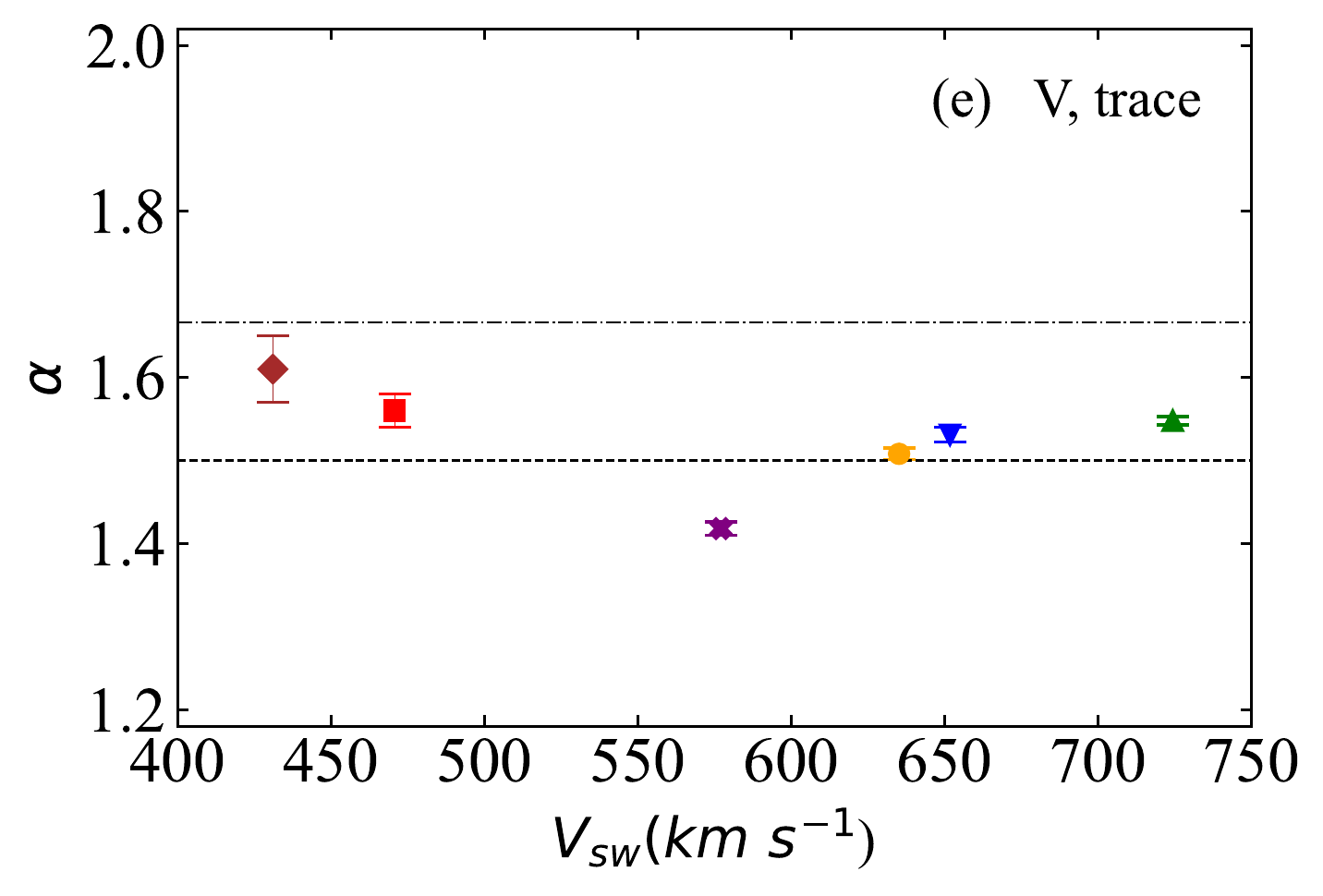}
    }
    \centerline{\includegraphics[width=0.49\textwidth,clip=]{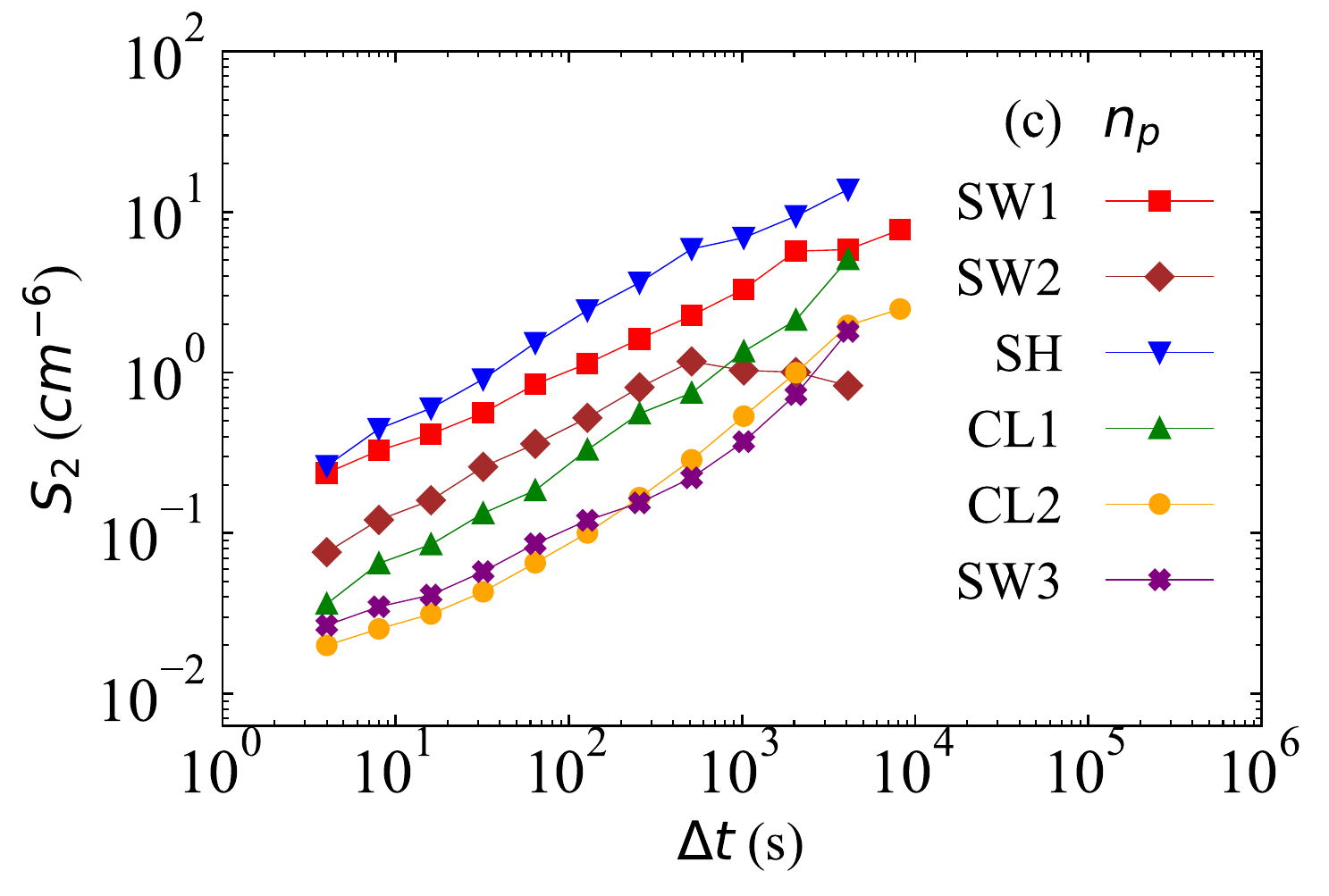} \includegraphics[width=0.49\textwidth,clip=]{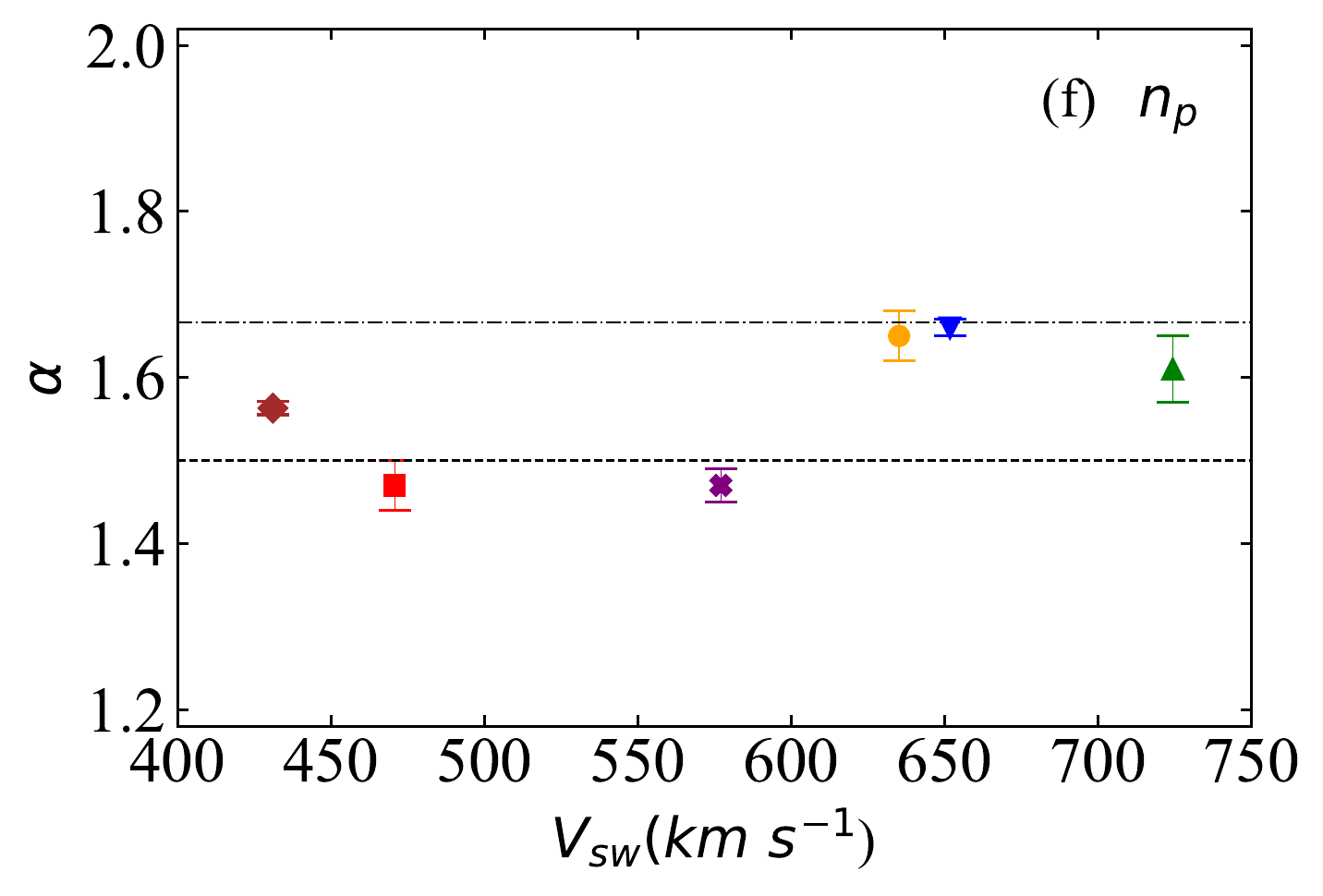} 
               }

\caption{{\it Panels (a)-(c):} second-order structure function $S_{2}$ versus different time scales $\deltat$ for magnetic field (a) and proton velocity (b) fluctuations averaged over the three components (indicated as ``trace''), and for proton density fluctuations (c), within all six selected intervals (see different colors and symbols in the legend). 
{\it Panels (d)-(f):} equivalent spectral indices $\alpha$ versus the mean solar wind speed, $\vsw$, displayed using the same colors and symbols as in panels (a)-(c). 
Dotted horizontal lines stand for the usual 5/3 and 3/2 values, corresponding to the K41 \citep{Kolmogorov1941} and IK64 spectra \citep{Iroshnikov1964,Kraichnan1965}, respectively. 
}
   \label{F-S2 panel}
   \end{figure}

The study of the scaling properties of the second-order structure function $S_{2}(\deltat)$ provides direct information on the turbulent energy spectra, for it is related to it through the power law dependence $S_{2}(\deltat) \propto \deltat ^{\alpha -1}$ \citep{Frisch1995}, where $\alpha$ is the spectral index. 
Computed values of $S_{2}$ plotted against different timescales for every sub-interval are shown in panels (a)-(c) of Figure \ref{F-S2 panel}. 
For the vector fields, the trace has been used. 
Results show a robust power-law dependence, with the possible exception of the density in the SW2 region (see panel (c)).
Power-law fits (not shown) were performed within timescales roughly between 10 and 1000 seconds, corresponding to the typical inertial range of solar wind plasmas of similar characteristics \citep[for comparison, see][]{BrunoCarbone2013,Kilpua2021,Sorriso-Valvo2021}, and several values of the equivalent spectral index $\alpha$ were thus obtained. 
The exponents were plotted versus the solar wind speed, $\vsw$, as shown in panels (d)-(f) of Figure \ref{F-S2 panel} (notice that, in the following, error bars will indicate the fitting parameters' standard deviation obtained from the covariance matrix after a standard $\chi^2$ optimization).

All the obtained spectral indices lie approximately between the usual values of 5/3 and 3/2, corresponding to the K41 and IK spectra, respectively, which supports the existence of a turbulent energy cascade. 
The largest values of $\alpha$ for the magnetic fluctuations was found in the sheath region (SH).
No apparent dependence between $\alpha$ and the mean solar wind velocity $\vsw$ was found, in accordance with previous observations \citep{Sorriso-Valvo2021}, nor between $\alpha$ and the large-scale cross helicity $|\sigmac|$ (not shown), the latter being evaluated near the large-scale end of the inertial range.

The intermittency of the turbulent fluctuations has been studied via the so-called flatness $F(\deltat)=S_{4}(\deltat)/S_{2}^{2}(\deltat)$, which provides an effective measure of the deviation from a Gaussian behaviour (for which $F=3$) of the $\deltat$-dependent distributions of the field increments.
Experimental evidence in fluid and plasma turbulence indeed consistently show the emergence of higher tails than those expected for Gaussian distributions, highlighting the generation of small-scale intermittent structures \citep{Frisch1995}. 
Furthermore, a power law of the kind $F(\deltat) \propto \deltat^{-\kappa}$, where $\kappa$ is the flatness scaling exponent, may be expected due to the scale invariance of the MHD equations within the turbulent inertial range.
Such negative power law is the consequence of the structure functions' anomalous scaling, namely of the deviation from the K41 prediction for their scaling exponents, $\zeta_q = hq$, in which case the flatness would be constant, resulting in $\kappa=0$ \citep{Frisch1995,BrunoCarbone2013,Carbone2014,Sorriso-Valvo2021}. 
The exponent $\kappa$ represents how effectively energy is transferred across scales, thus being greater values of $\kappa$ related to a faster formation of small-scale turbulent structures, or to stronger intermittency.

\begin{figure}    
   \centerline{
    \includegraphics[width=0.49\textwidth,clip=]{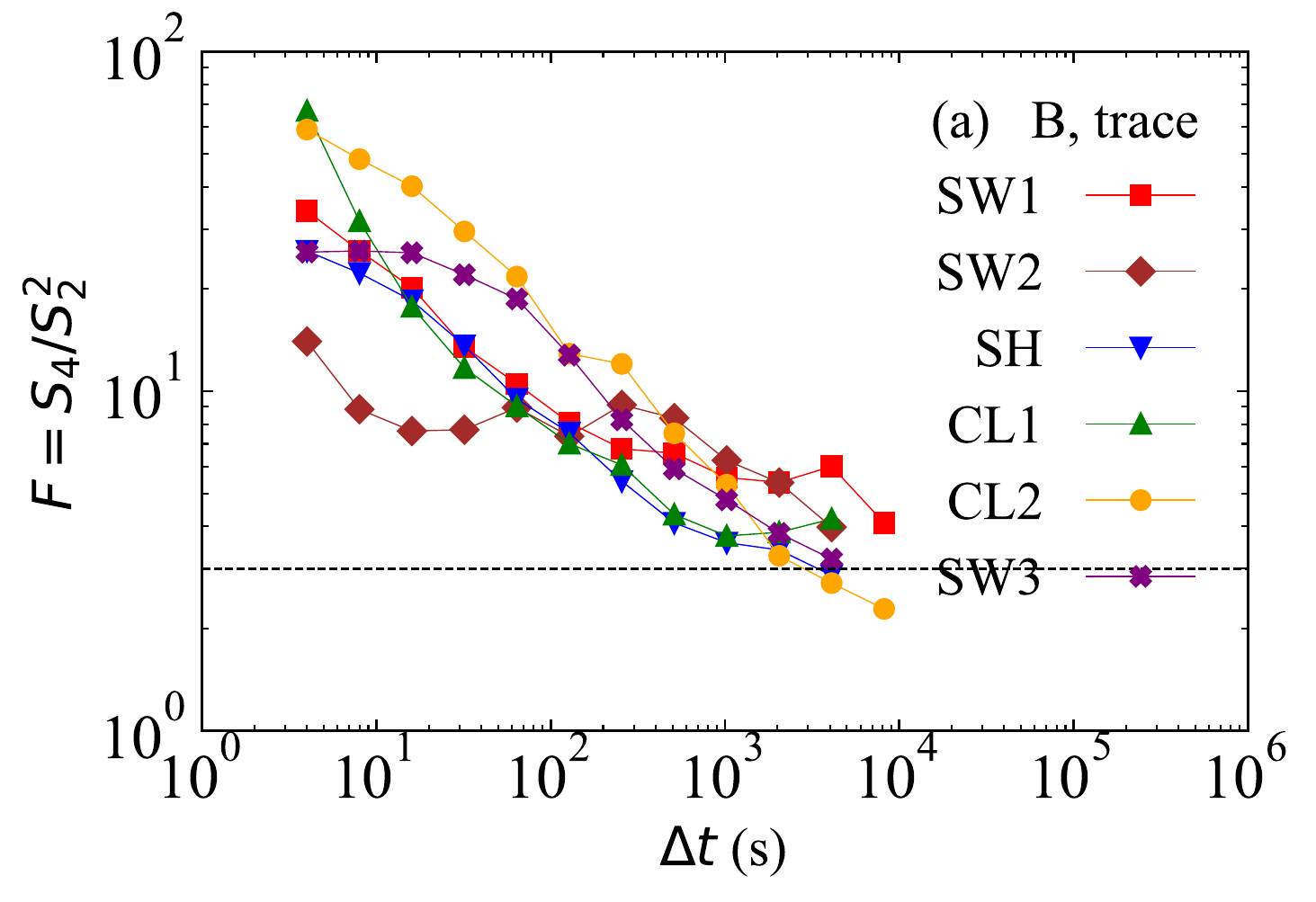} \includegraphics[width=0.49\textwidth,clip=]{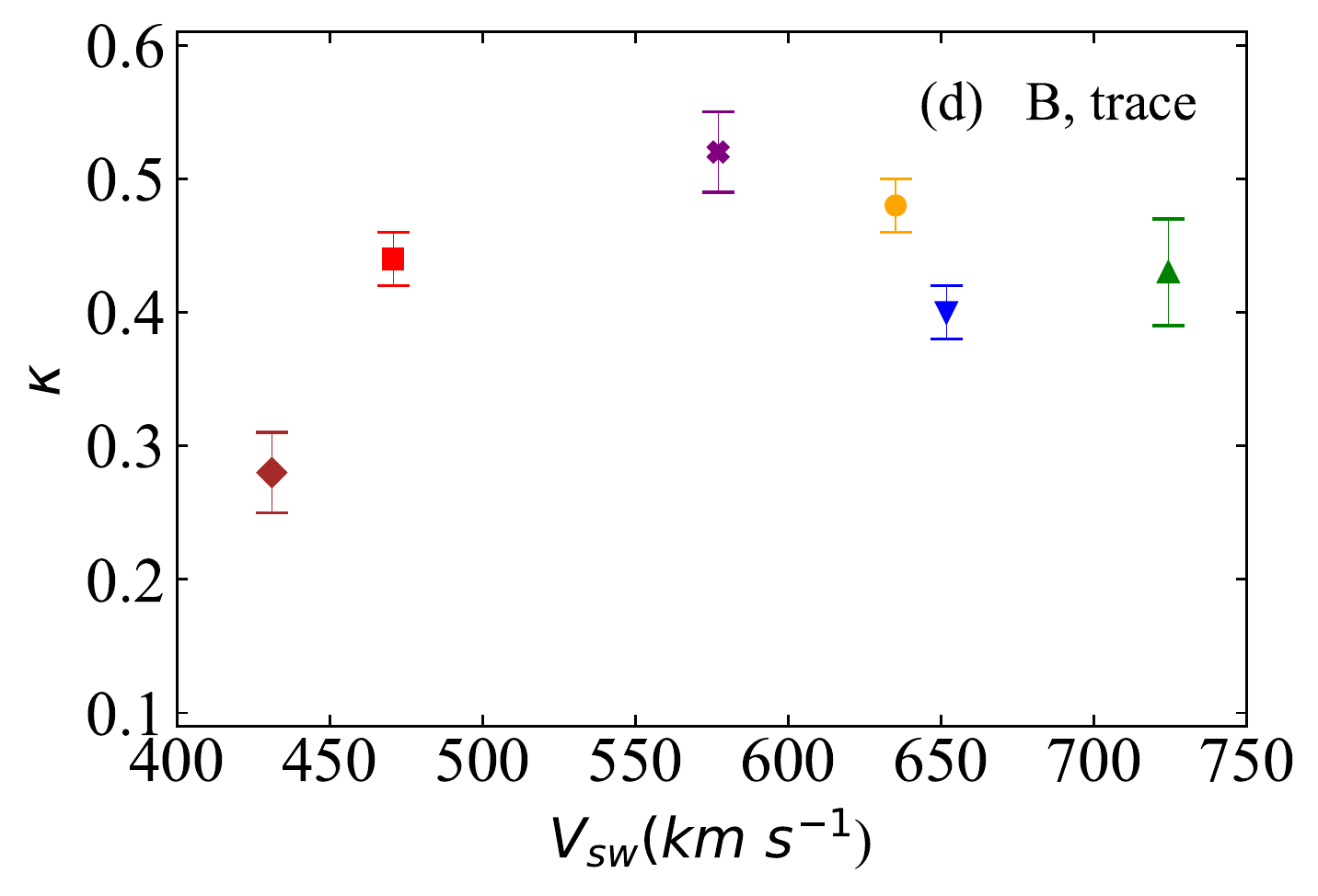} 
    }

    \centerline{    \includegraphics[width=0.49\textwidth,clip=]{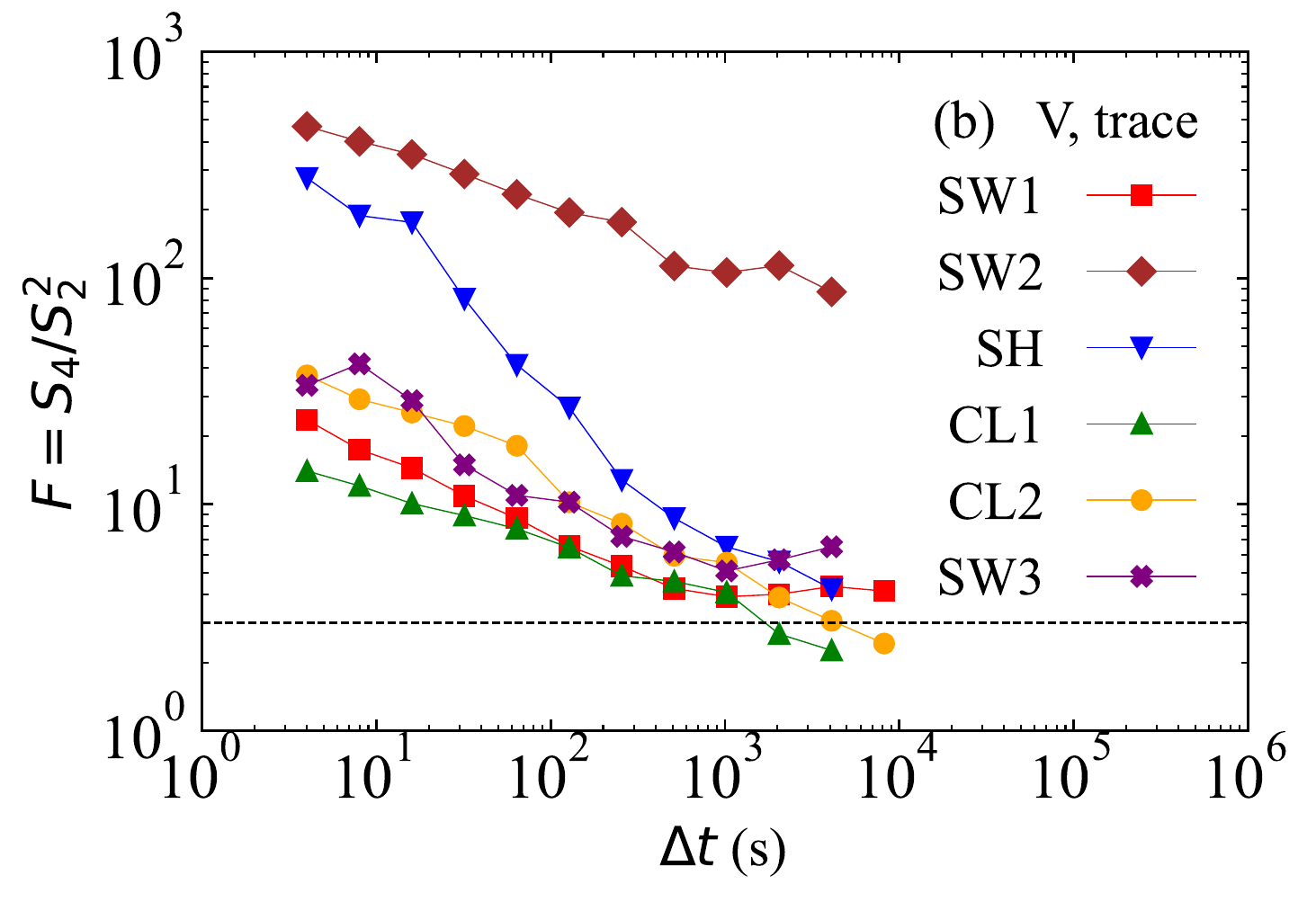}  \includegraphics[width=0.49\textwidth,clip=]{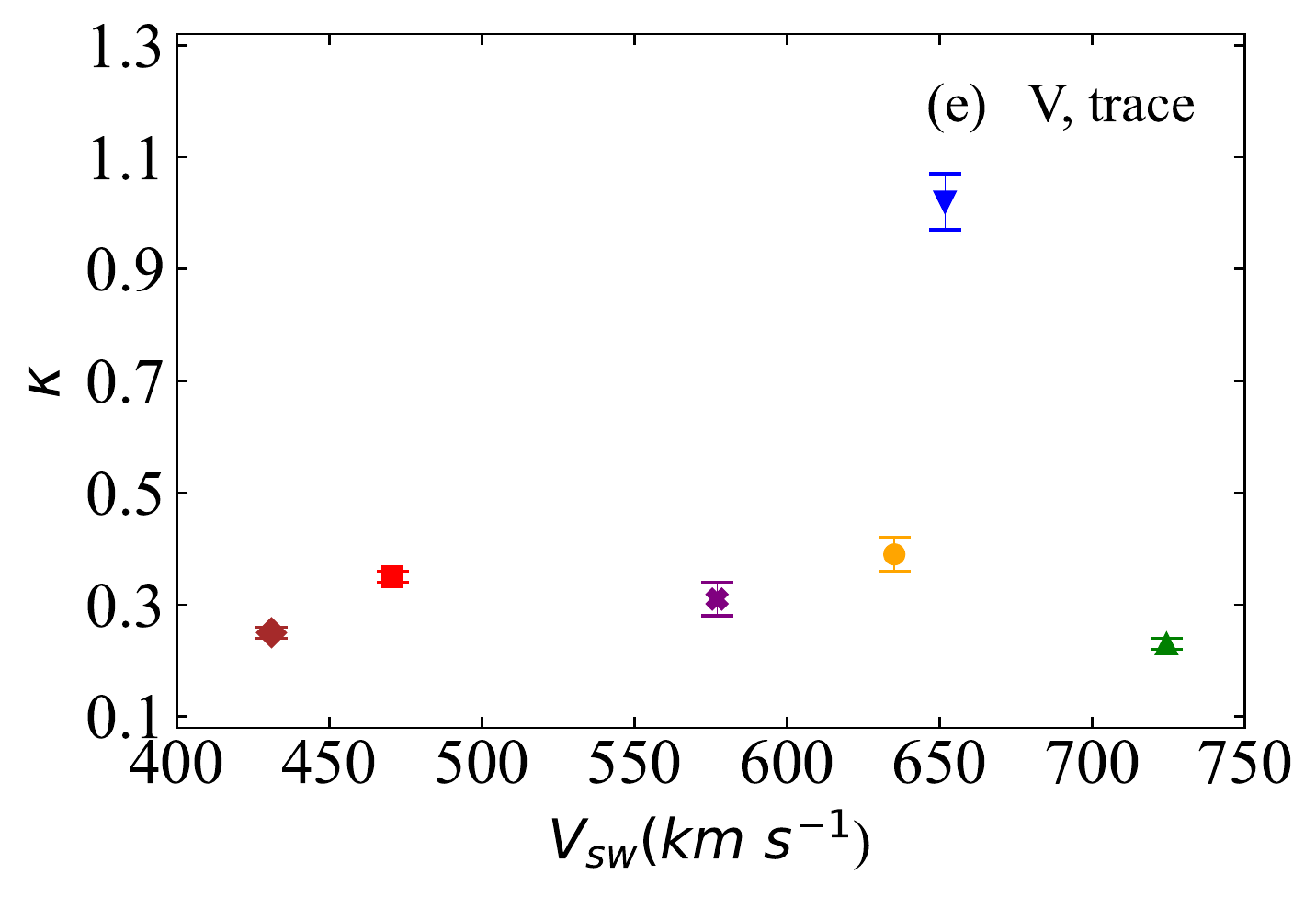}
    }
    
    \centerline{    \includegraphics[width=0.49\textwidth,clip=]{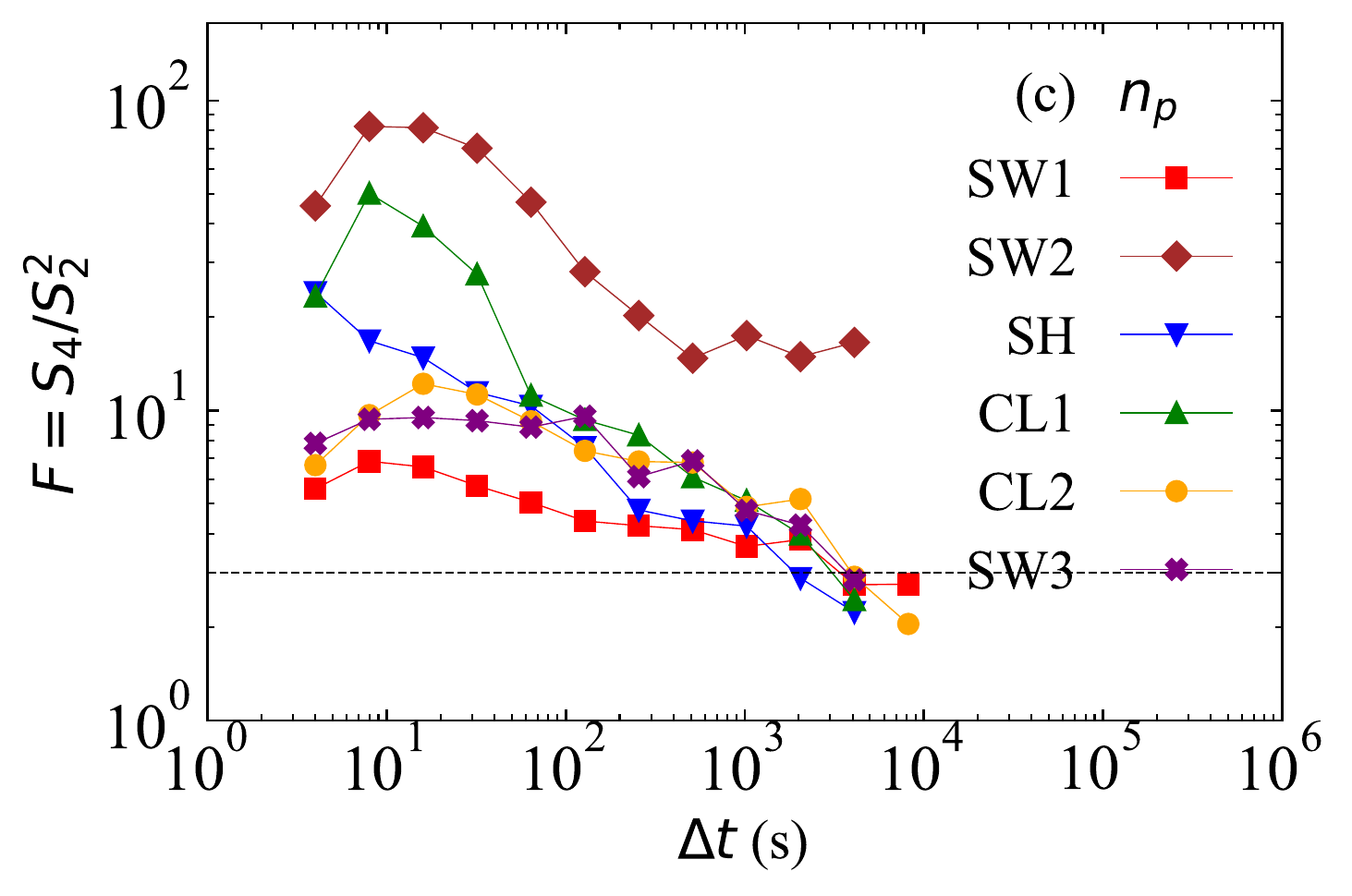} 
    \includegraphics[width=0.49\textwidth,clip=]{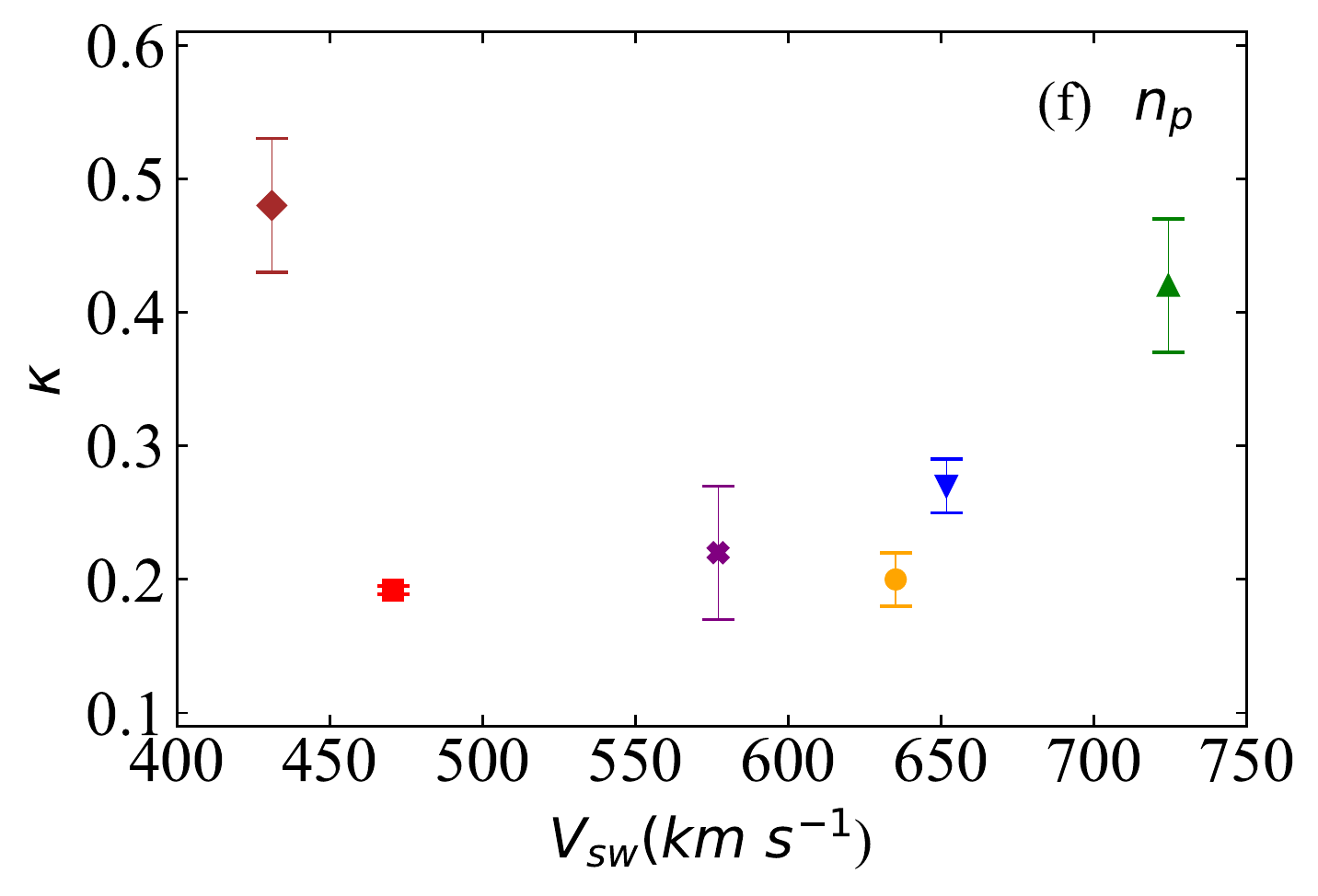} 
              }
              
\caption{{\it Panels (a)-(c):} Flatness $F=S_{4}/S^{2}_{2}$ versus different time scales $\deltat$ for magnetic field (a) and proton velocity (b) fluctuations averaged over the three components (indicated as ``trace''), and for proton density fluctuations (c), within all six selected intervals (see different colors and symbols in the legend). Dotted, horizontal lines correspond to the Gaussian value, $F=3$. 
{\it Panels (d)-(f):} flatness scaling exponents, $\kappa$, versus the mean solar wind speed $\vsw$, displayed using the same colors and symbols as in panels (a)-(c). 
Note the different y-axis range in panel (e). 
}
   \label{F-F panel}
   \end{figure}


Plotted values of $F$ against different $\deltat$ are shown in panels (a)-(c) of Figure \ref{F-F panel} for all the selected intervals.
Results show that the flatness behaves mostly as a negative power law within the inertial range, indicating intermittency. 
Several values of $\kappa$ were obtained via power-law fits (not shown), being their values plotted versus $\vsw$ in panels (d)-(f) of Figure \ref{F-F panel}. 
Their values lie between the typically observed range for space plasma measurements, 0.1---0.5 \citep{Sorriso-Valvo2018,Sorriso-Valvo2021,Hernandez2021,Quijia2021}. 
As often found in space plasmas, magnetic field intermittency is quite consistently higher than for velocity and density \citep{Sorriso-Valvo1999}. 
One of the most relevant features of Figure \ref{F-F panel}, easily seen in panel (e), is the exceptionally large exponent obtained for the velocity in the sheath region (blue), which is much higher than in the other regions. Such large value indicates an enhanced presence of velocity fluctuations, generated by the ICME shock. 
This result resembles that obtained by \citet{Sorriso-Valvo2021} for the proton density fluctuations and, as we shall see in the following Section \ref{S-Third-order moment scaling law}, it is linked with a much higher mean turbulent energy transfer rate within the ICME sheath. 
As in the case of the spectral exponents, no clear correlations were found between the flatness scaling exponents and $\vsw$, nor between them and the cross-helicity $\sigma_{c}$. 
An exception is perhaps the moderate correlation observed between the intermittency exponent and the solar wind speed for the proton density (panel (f)), showing that the most compressed regions (i.e., the sheath and the first cloud section, see also the cross-helicity in the top panel of Figure \ref{F-epsilon panel}) have also enhanced density intermittency.

The structure-function analysis revealed the presence of well developed turbulence in all of the examined sub-intervals, with the possible exception of the density in SW2, where the scaling does not extend to the typical two or more decades as for the other cases. 
The turbulence is compatible with Kolmogorov or Iroshnikov-Kraichnan phenomenology, and intermittency is moderate to strong, being exceptionally strong for the velocity in the ICME sheath region.

\section{Third-order moment scaling law}
     \label{S-Third-order moment scaling law}

\citet{Politano1998} derived a relevant exact result for MHD turbulence that replicates the fundamental Kolmogorov's 4/5 law for neutral flows \citep{Kolmogorov1941} and that has been observed in solar wind plasmas for more than a decade \citep{Sorriso-Valvo2007}. 
Known as the Politano-Pouquet law (PP law, hereafter), it can be stated as follows:
\begin{equation}  \label{Eq-PP law}
Y(\Delta t):=\left\langle\Delta v_L\left(|\Delta \bv|^2+|\Delta \bb|^2\right)-2 \Delta b_L(\Delta \bv \cdot \Delta \bb)\right\rangle=\frac{4}{3} \varepsilon V_{s w} \Delta t \, \,.
\end{equation} 
This law involves mixed third-order structure functions (left hand side in Equation \ref{Eq-PP law}) and arises as a direct consequence of the incompressible MHD equations, once statistical homogeneity, stationarity, a high Reynolds number and local isotropy are assumed \citep[see][and references therein]{Marino2023}. 
The PP law describes the turbulent cascade, defining rigorously the inertial range, and providing information on the mean energy transfer rate ($\varepsilon$ in Equation \ref{Eq-PP law}) across scales. 
$\Delta v_L$ and $\Delta b_L$ stand for longitudinal timescale-dependent increments (denoted with $L$, in this case referring to the direction of the bulk solar wind flow, namely from the Sun to the Earth), of the plasma and Alfv\'enic velocities, respectively, via $\Delta \phi_L=\phi_L(t+\Delta t)-\phi_L(t)$; brackets stand for time averages over the samples. 
The linear relation in Equation (\ref{Eq-PP law}) provides a valuable tool to estimate the energy transfer rate of the turbulent cascade, $\varepsilon$, directly from the measurements. 
In addition, the sign of the energy transfer rate is associated with the direction of the energy cross-scale flux. A positive transfer rate indicates a direct cascade, with the energy flowing predominantly from larger to smaller scales. On the other hand, negative $\varepsilon$ could be associated with an inverse cascade, where the energy is mostly transferred from smaller to larger scales \citep{Politano1998,Marino2012,Smith2009}. This can be the case, for example, because of some scale-local energy input. 
However, the transfer rate sign can also flip due to local inhomogeneity or anisotropy \citep{Stawarz2011,Smith2009,Verdini2015b,Coburn2015,Hernandez2021,Marino2022}, so that the relation between cascade sign and direction is still an open question \citep{Marino2023}.

\begin{figure}    
   \centerline{\hspace*{0.001\textwidth}
               \includegraphics[width=0.5\textwidth,clip=]{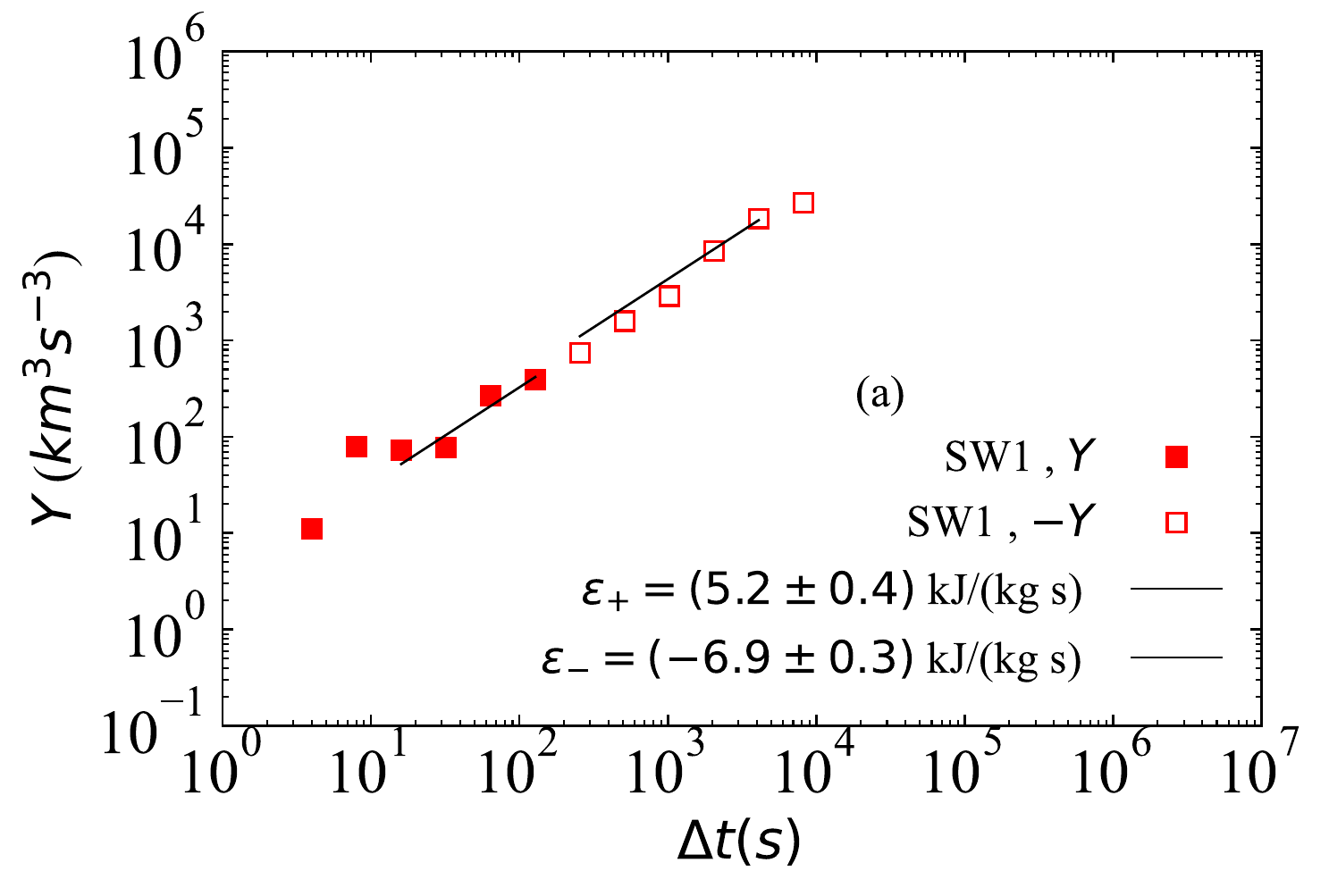} 
            
               \hspace*{-0.001\textwidth}
               \includegraphics[width=0.5\textwidth,clip=]{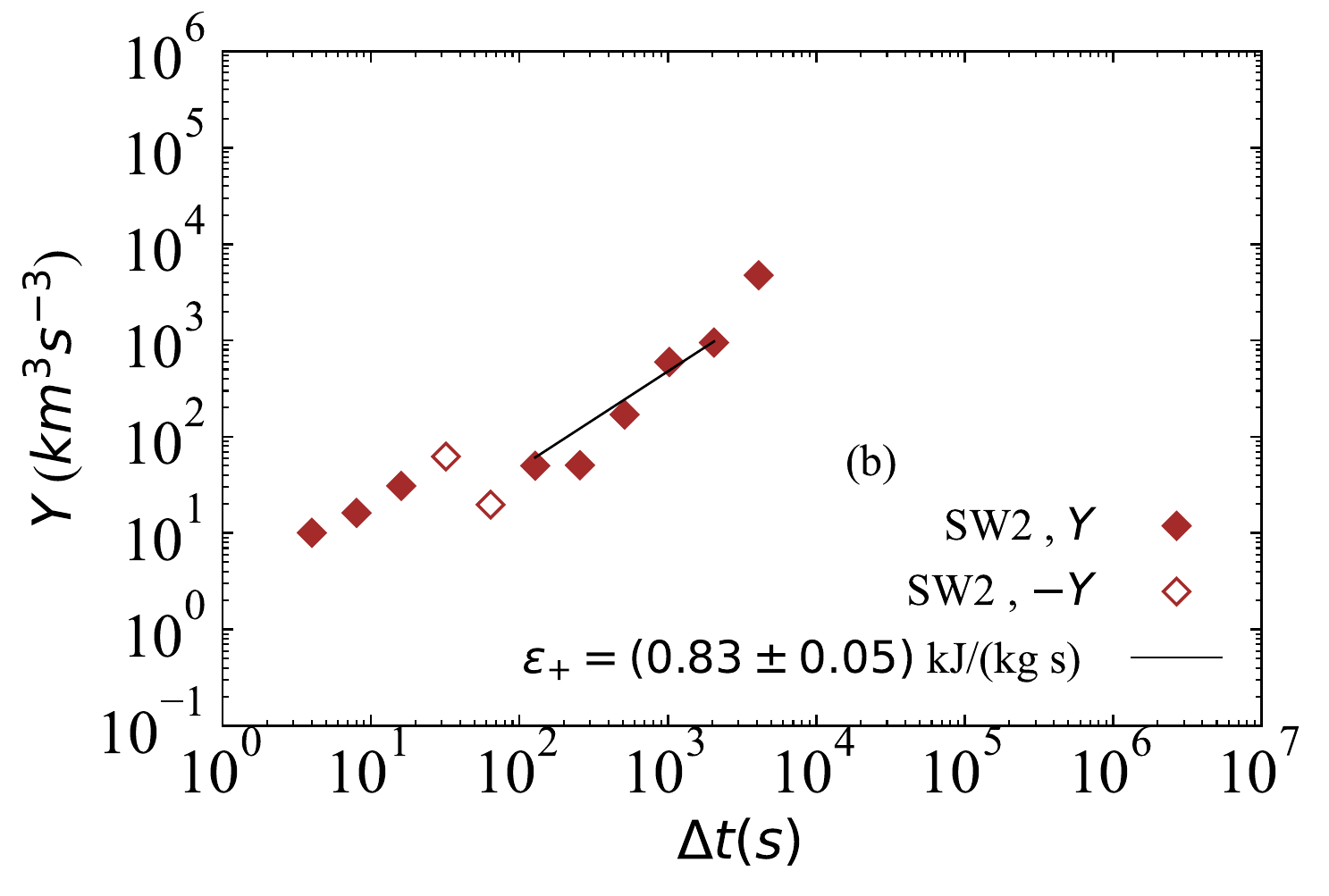} 
              }
   \centerline{\hspace*{0\textwidth}

               \includegraphics[width=0.5\textwidth,clip=]{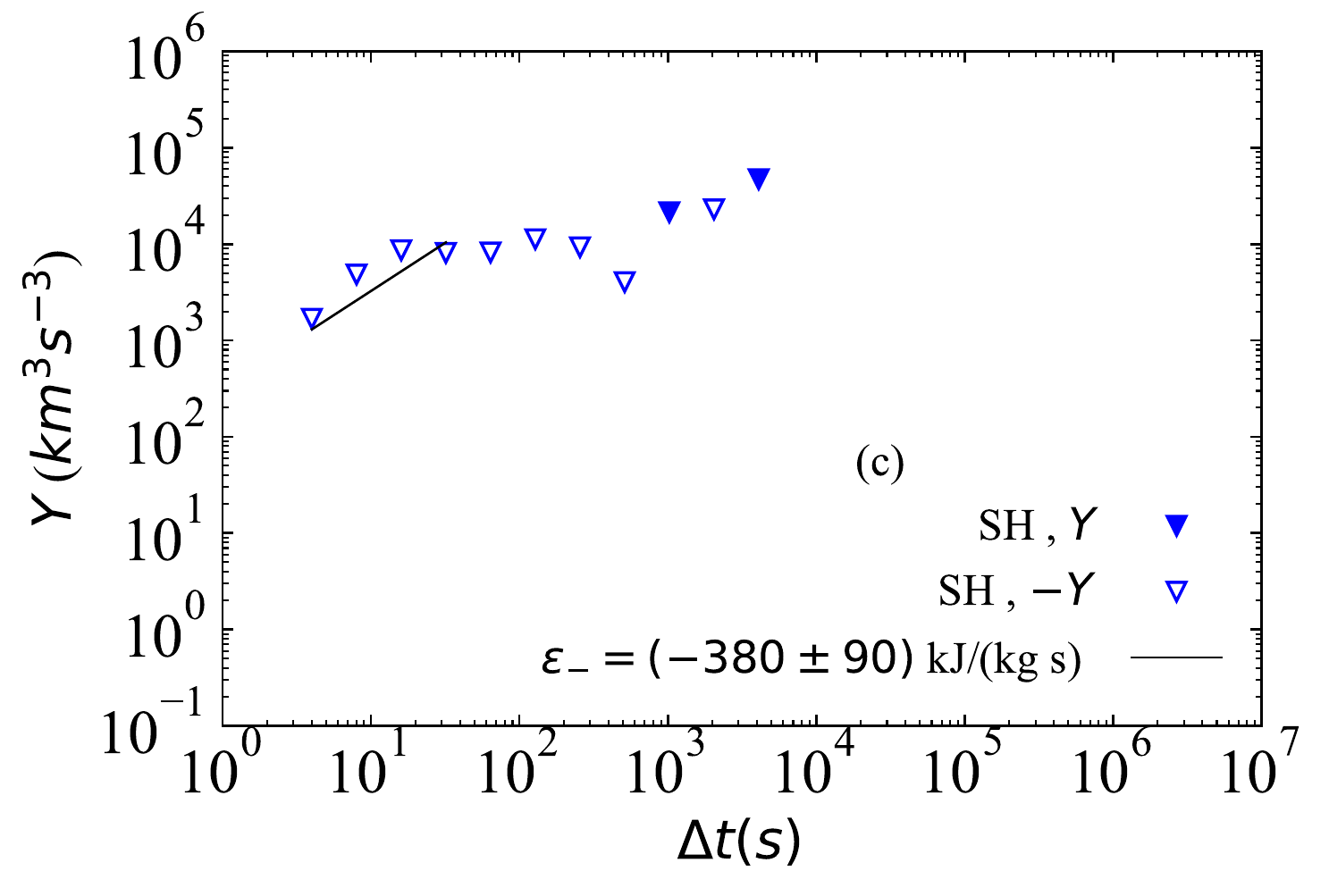} 

               \hspace*{-0.001\textwidth}
               \includegraphics[width=0.5\textwidth,clip=]{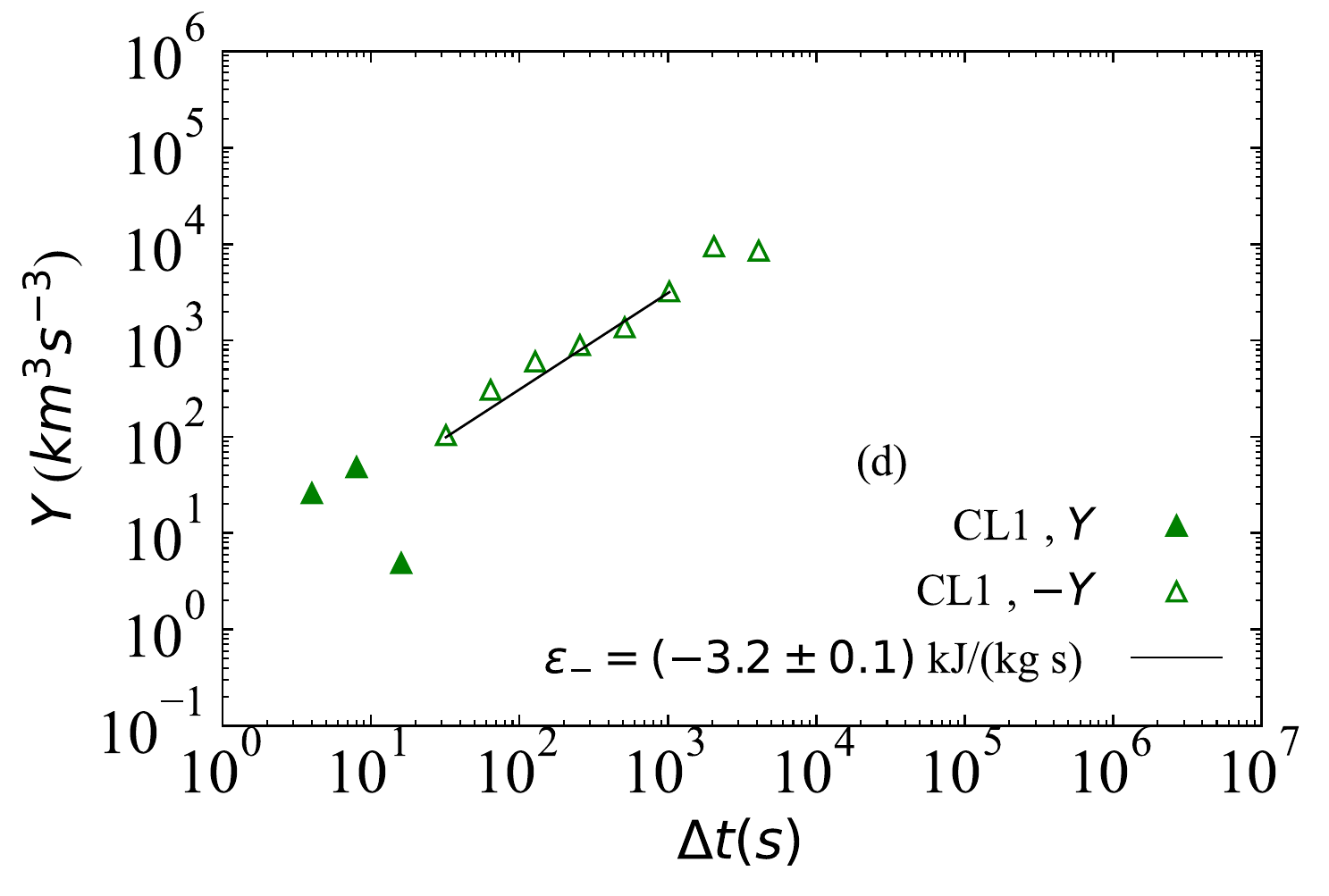} 
               }

    \centerline{\hspace*{0.001\textwidth}
               \includegraphics[width=0.5\textwidth,clip=]{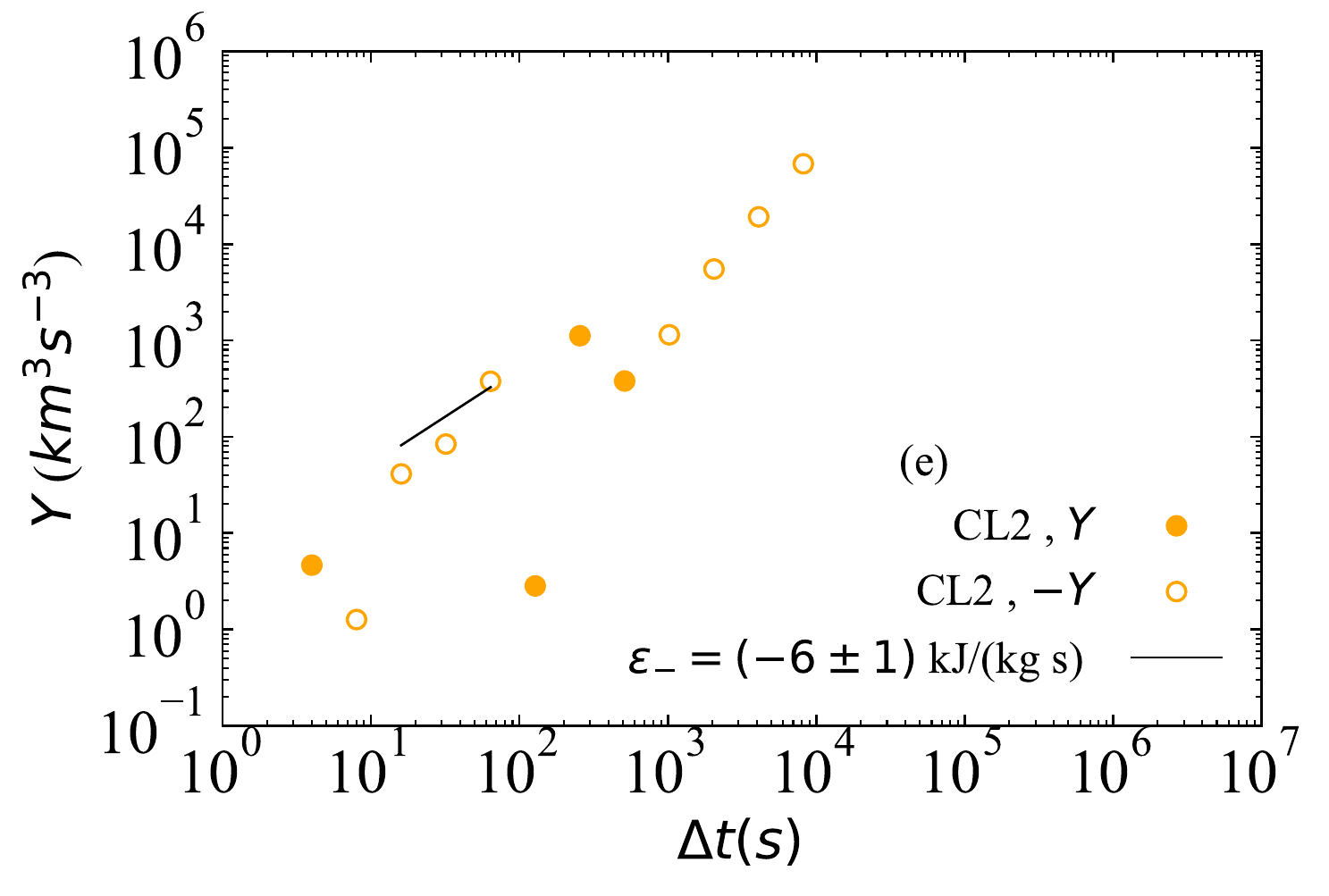} 

               \hspace*{-0.001\textwidth}
               \includegraphics[width=0.5\textwidth,clip=]{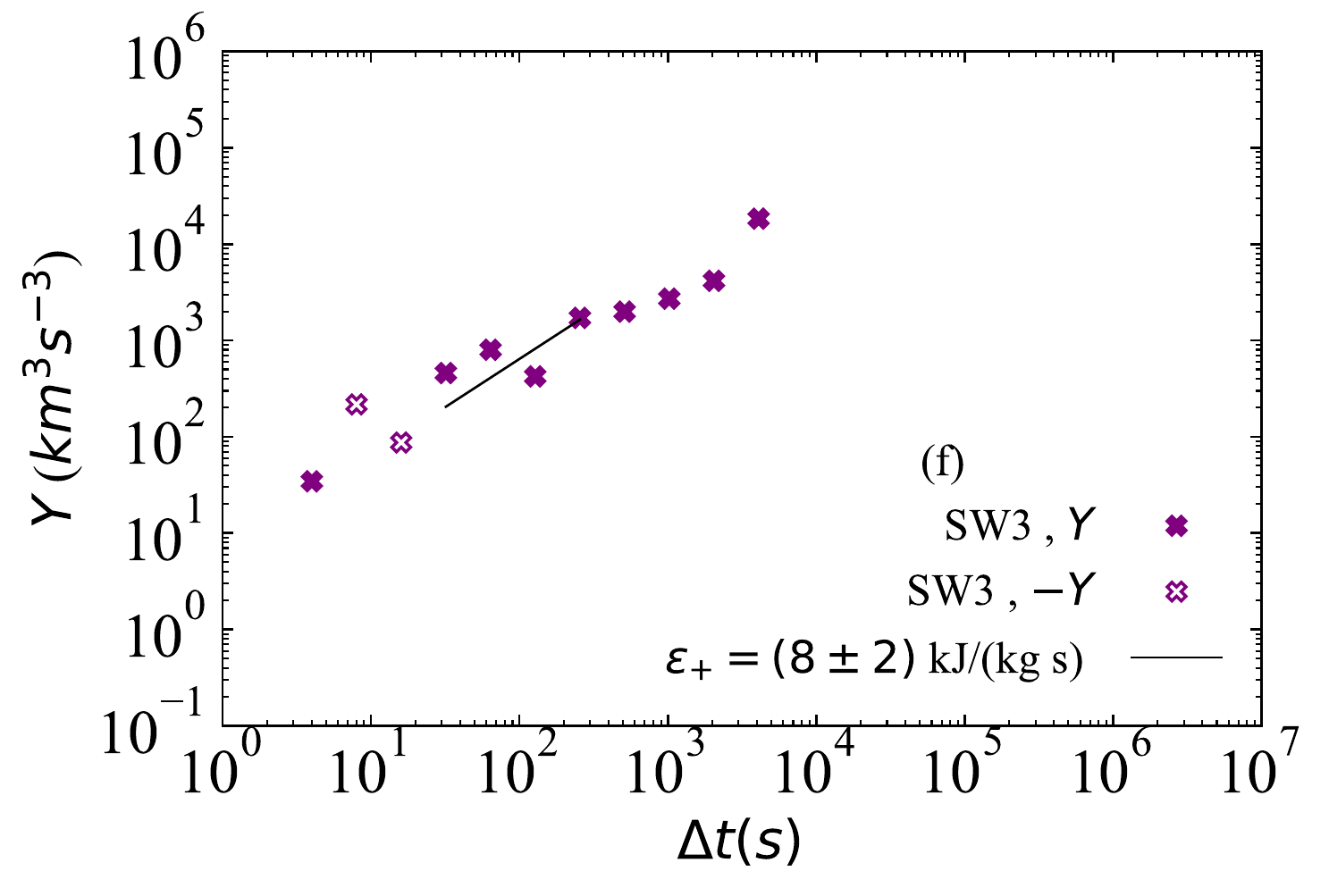}
               
              }
              
\caption{{\it Panels (a)-(f):} mixed third-order moment, $Y(\Delta t)$ (Equation \ref{Eq-PP law}), for the six selected intervals (different colors and symbols). Colour-filled (empty) markers indicate positive (negative) values. 
Linear fits (solid lines) were performed in intervals corresponding to the inertial range. The fitted values of $\varepsilon$ are shown in each panel (the error being the standard deviation from the linear fits). 
Positive and negative energy transfer rates are labeled as $\varepsilon_{+}$ and $\varepsilon_{-}$, respectively. 
}  
   \label{F-Yaglom panel}
   \end{figure}

We shall proceed now to study the PP law within this 12-14 September 2014 interplanetary coronal mass ejection. 
Results are shown in Figure \ref{F-Yaglom panel}, where panels (a)-(f) display computed values of $Y$ in equation \ref{Eq-PP law} plotted against different timescales $\deltat$, for all six selected intervals. 
Colour-filled (empty) markers indicate positive (negative) values of Y; the latter have been reverted so as to be represented through base-10 logarithmic axis. 
The performed linear fits, done in order to obtain values of the mean energy transfer rates, are also shown, being the values of $\varepsilon$ corresponding to positive or negative $Y$ (denoted as $\varepsilon_{+}$ and $\varepsilon_{-}$, respectively) displayed on the (a)-(f) panels, with their corresponding fitting uncertainties. Notice that panel (a) provided two estimated $\varepsilon$ values, corresponding to positive or negative $Y_{SW1}$ separately.

The SW1 and CL1 panels clearly show the linear behaviour expected from the PP law. On the other hand, the linear scaling is not as good in the remaining sub-intervals. However, in those cases, where the lack of statistical convergence affects the regularity of the scaling, it is still possible to obtain reasonable estimates of the mean energy transfer rate. 
Interestingly, the SW1 interval, shown in panel (a), reveals a very clear and clean sign reversal at scales of a few minutes. Similar reversals were observed before in the solar wind \citep{Sorriso-Valvo2007}. While in some cases those were ascribed to the presence of anisotropy effects \citep{Stawarz2011} or to the switch in the dominance of inward or outward Elsasser modes \citep{Coburn2015}, in other cases sign reversal were found across intervals characterized by the abundant presence of switchbacks or other structures of size comparable with the scale of the sign flip \citep{Hernandez2021}. 
This might suggest, at least in some cases, that the sign flip is related to the presence of an energy injection at such scale, which might be feeding simultaneously a direct and an inverse cascade.
Understanding the actual significance of sign reversals, and more generally of the observed sign of the cascade, is unfortunately more complex than suggested by the original PP theoretical result, and deserves in-depth studies that are outside of the scope of this paper \citep{Marino2022,Marino2023}.

\begin{figure}    
 \centerline{
                   \includegraphics[width=0.45\textwidth,clip=]{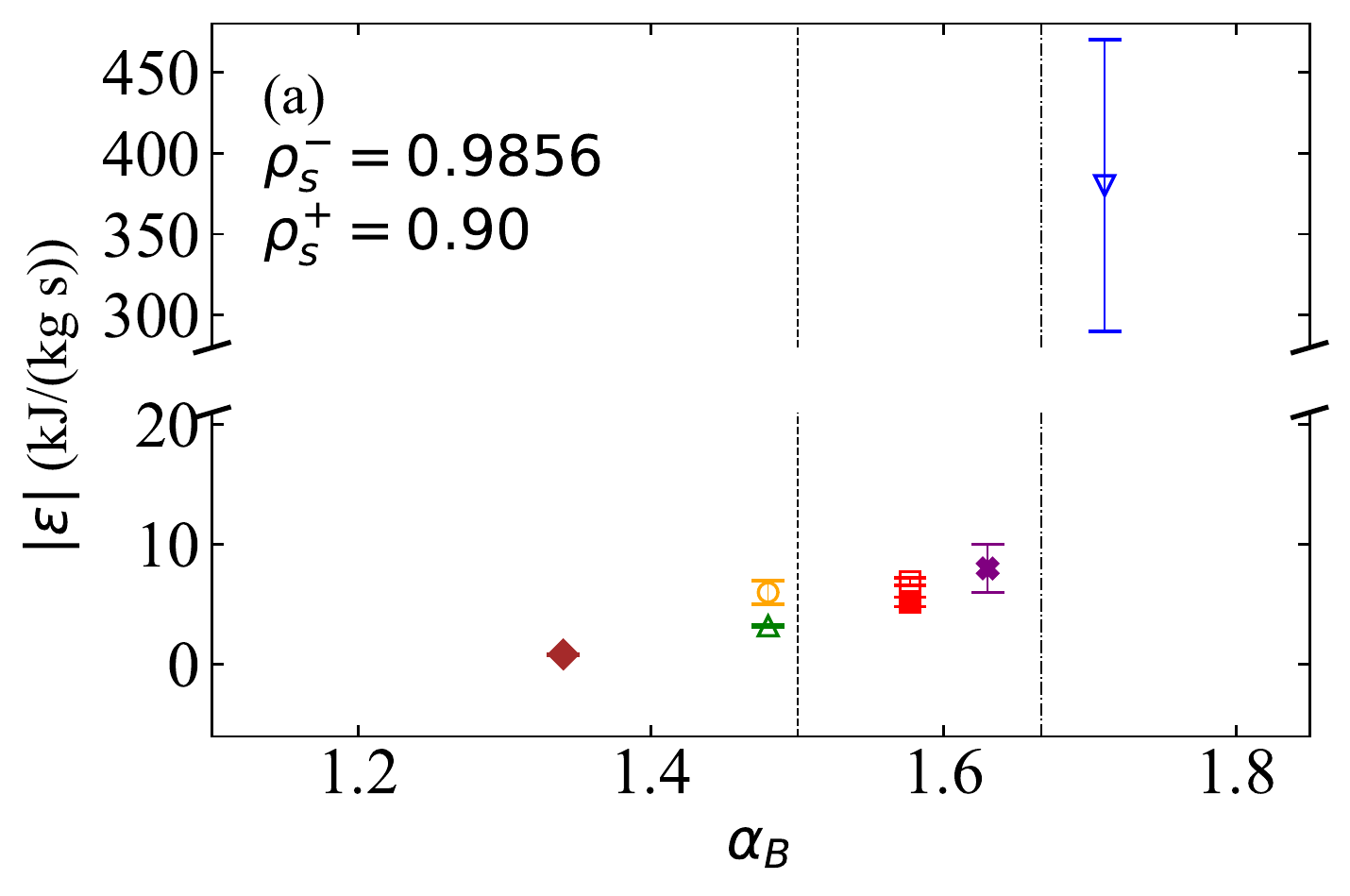}

                   \includegraphics[width=0.45\textwidth,clip=]{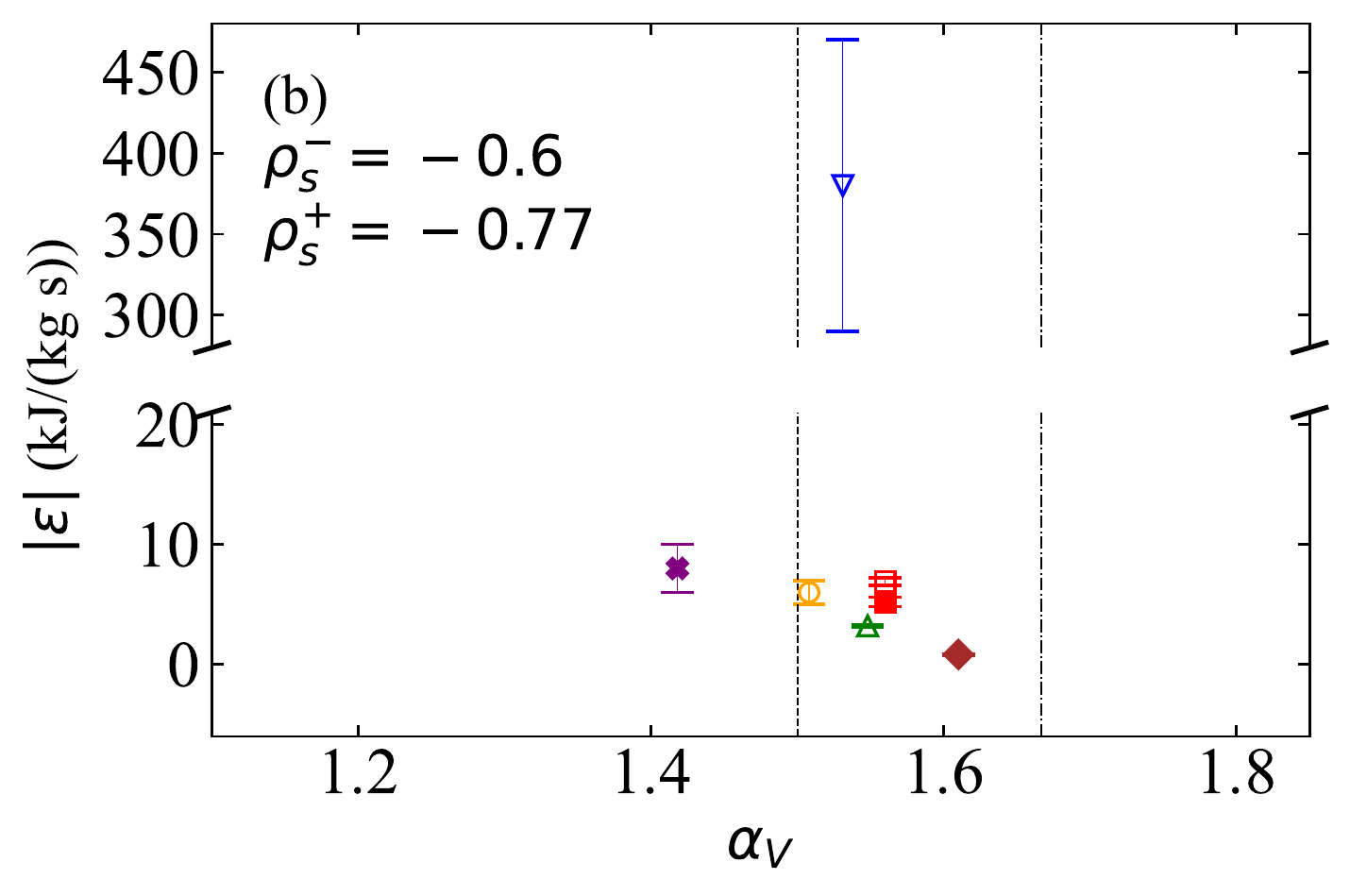}
              }

       \centerline{       \includegraphics[width=0.45\textwidth,clip=]{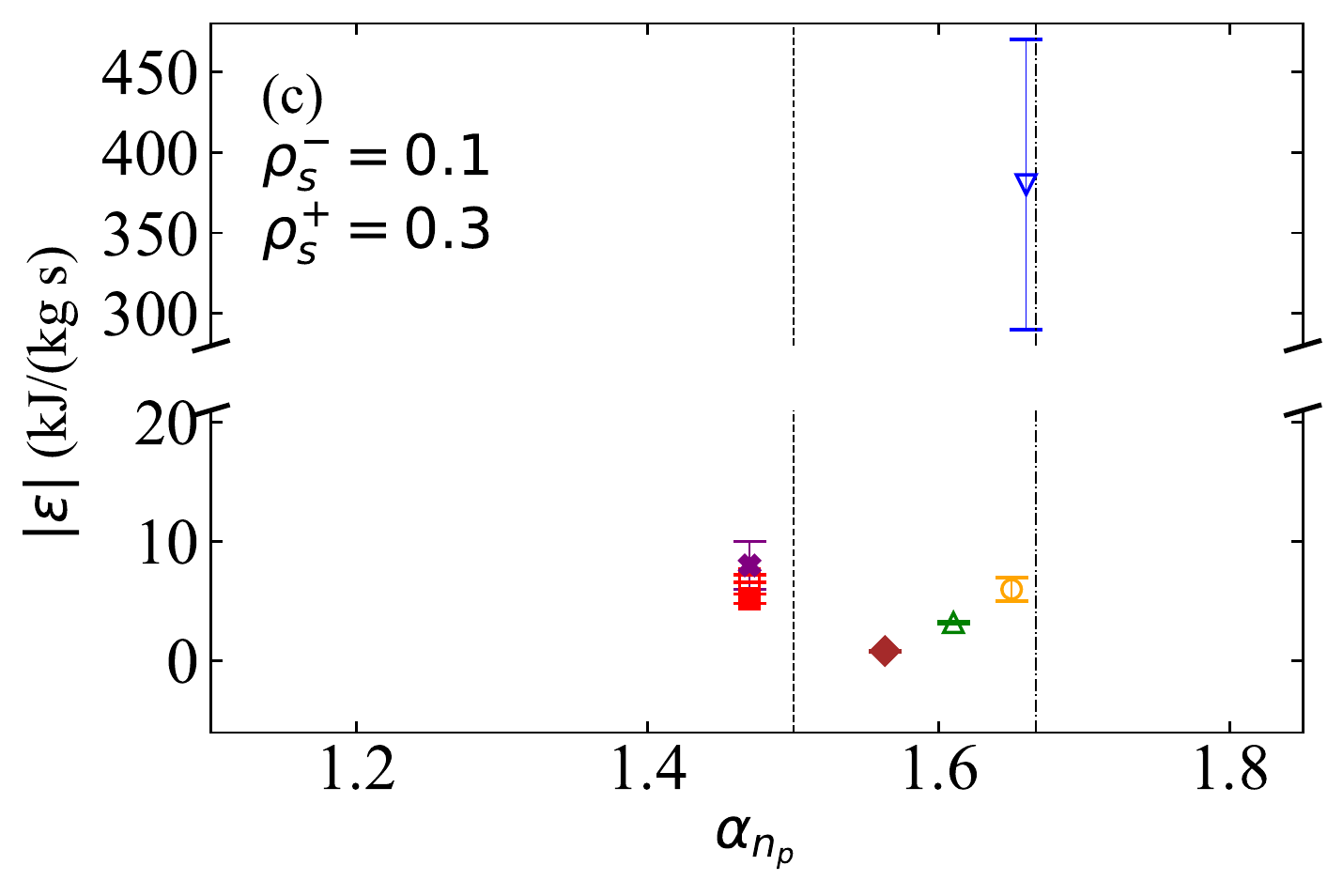}
        \includegraphics[width=0.45\textwidth,clip=]{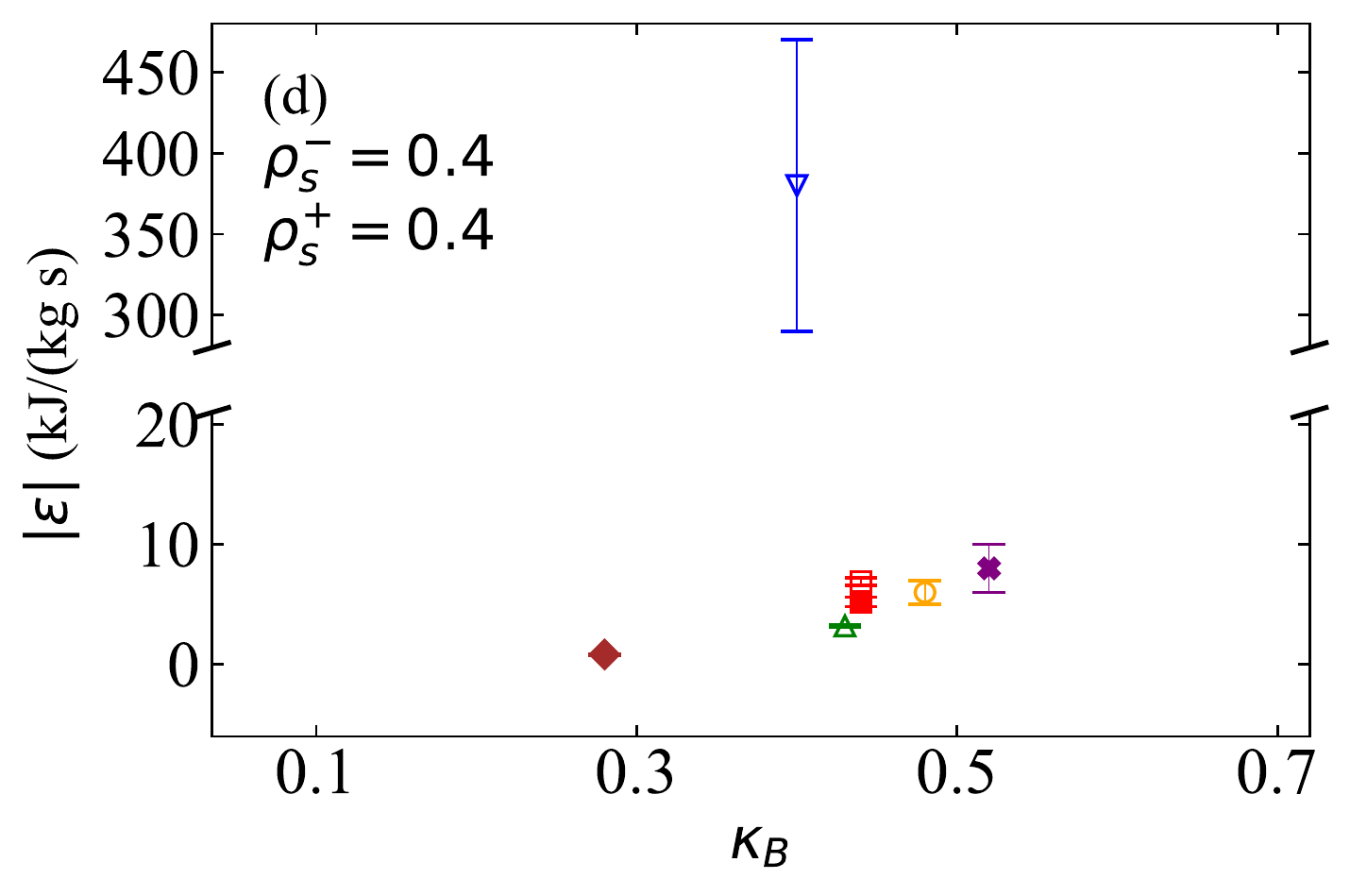}
              }
    \centerline{
                   \includegraphics[width=0.45\textwidth,clip=]{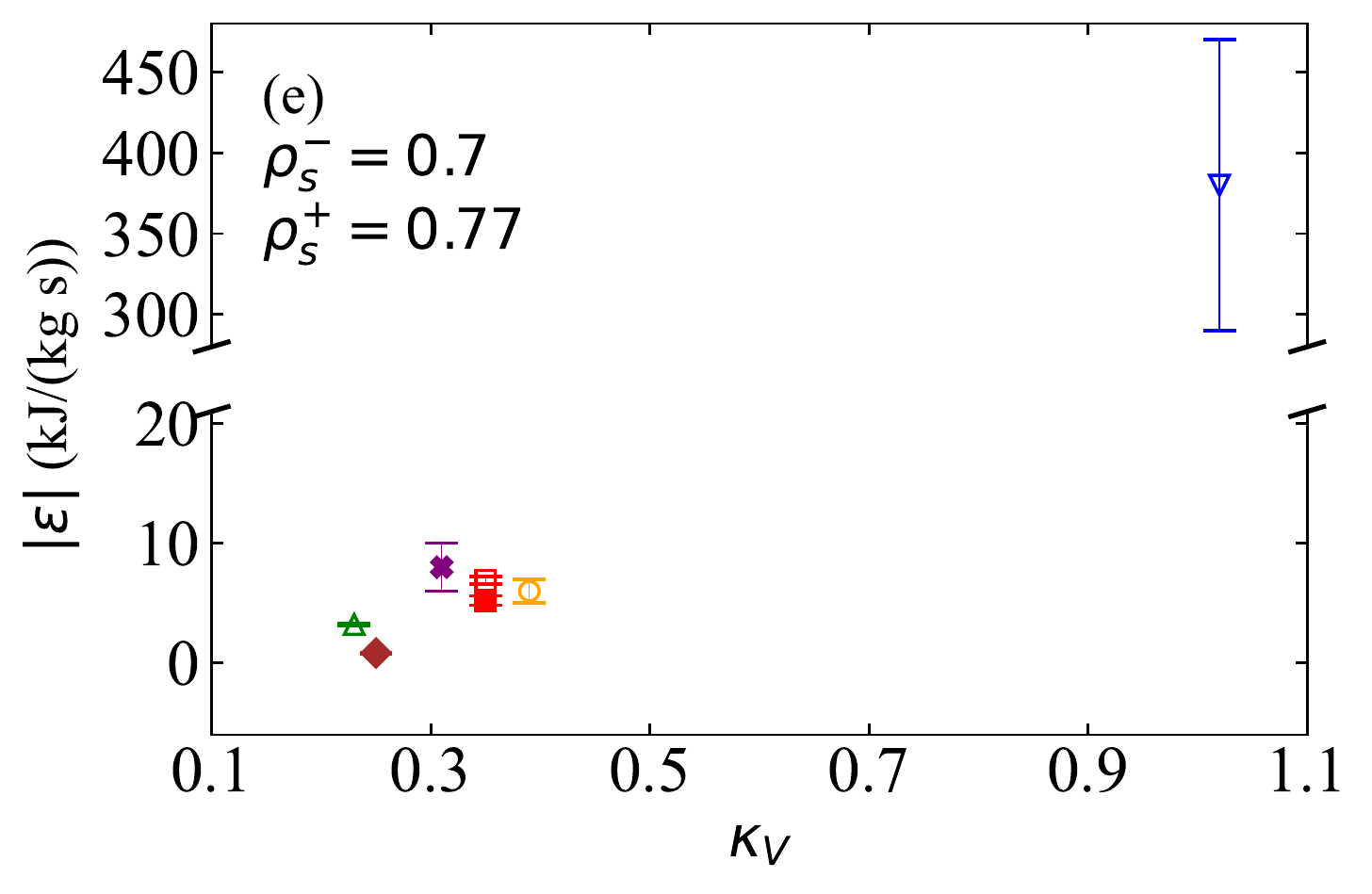}
        
                   \includegraphics[width=0.45\textwidth,clip=]{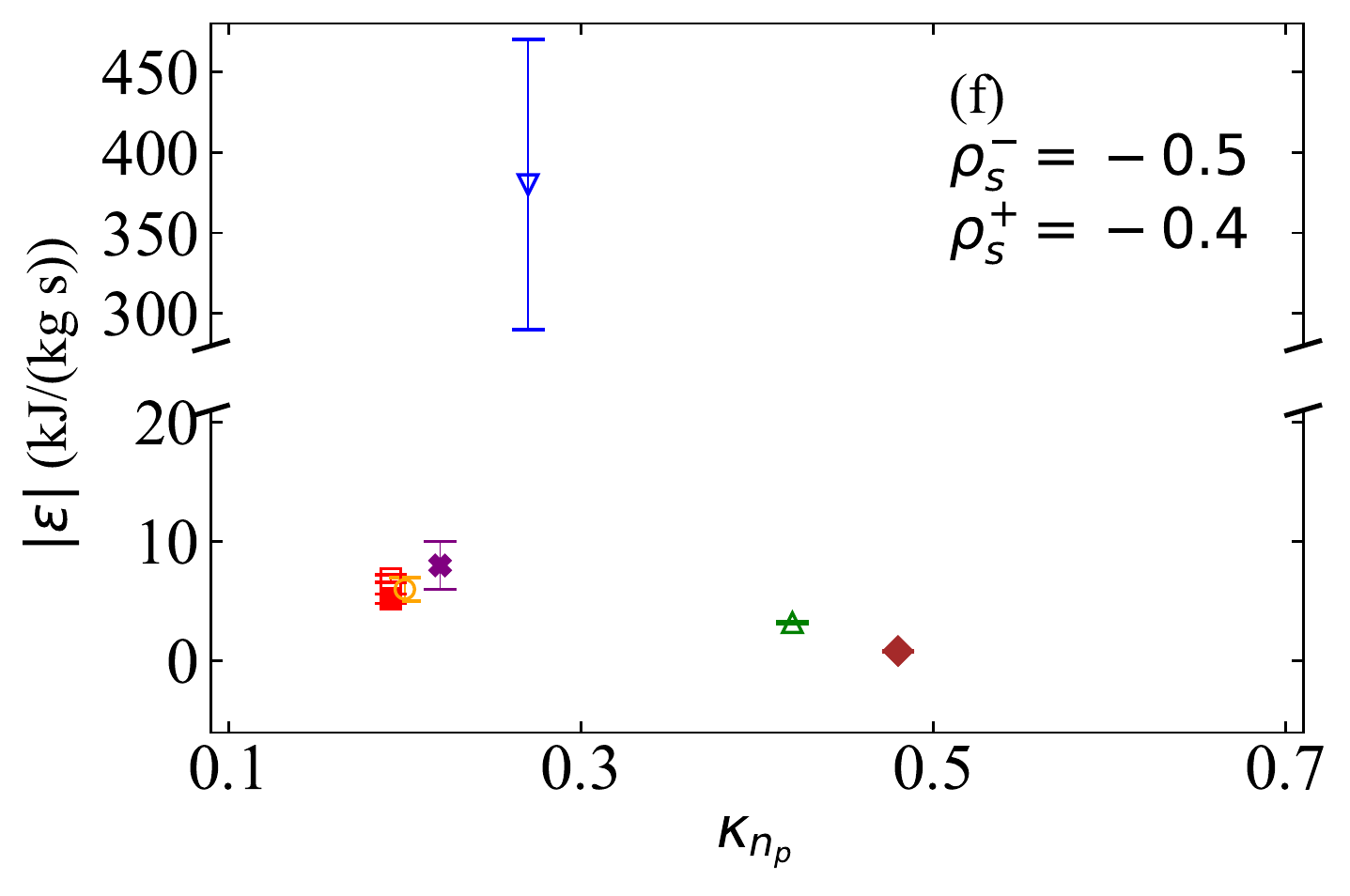}
             }
    \centerline{\hspace*{0.001\textwidth}
\hspace{4pt}\includegraphics[width=0.45\textwidth,clip=]{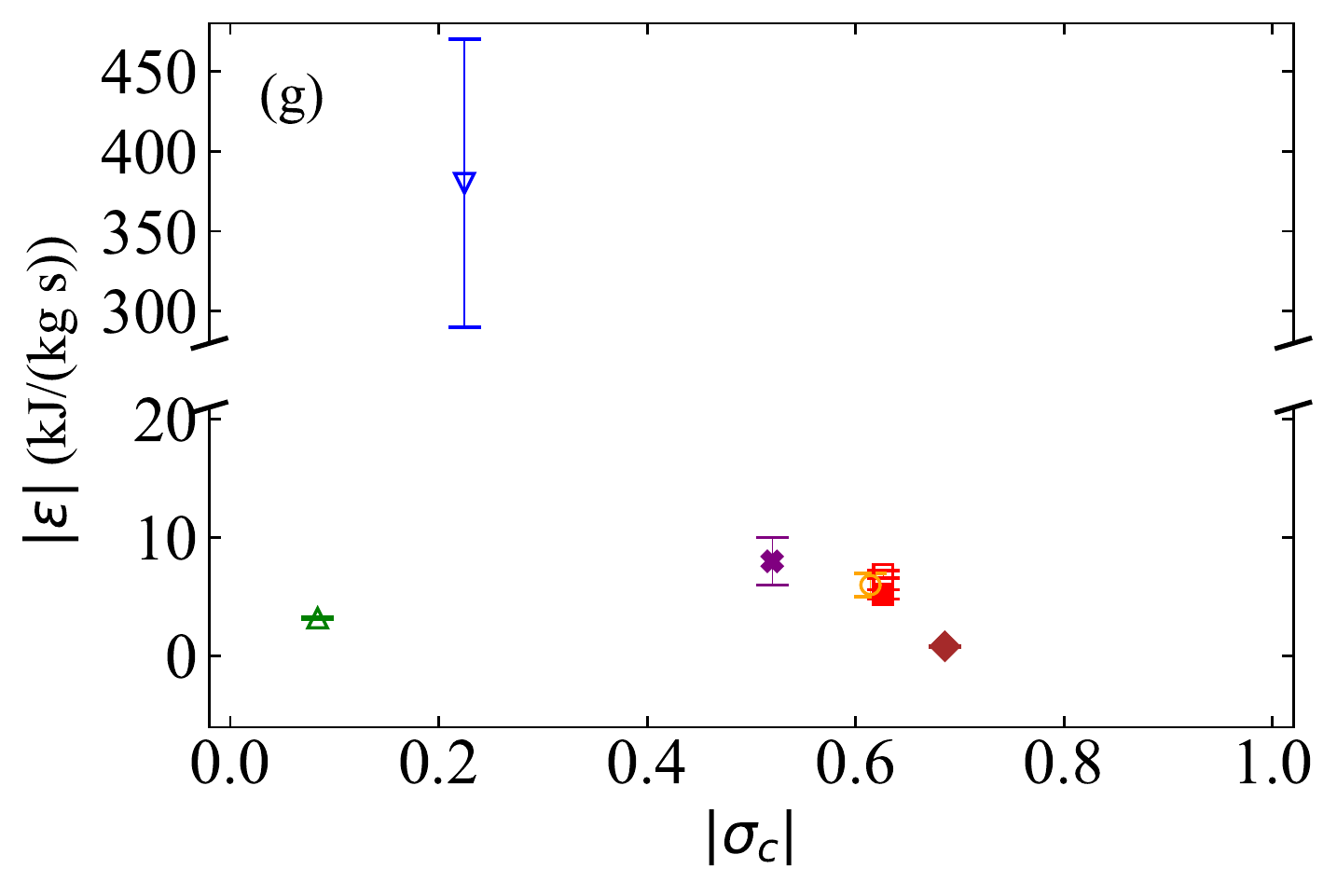}
\hspace{4pt}\includegraphics[width=0.45\textwidth,clip=]{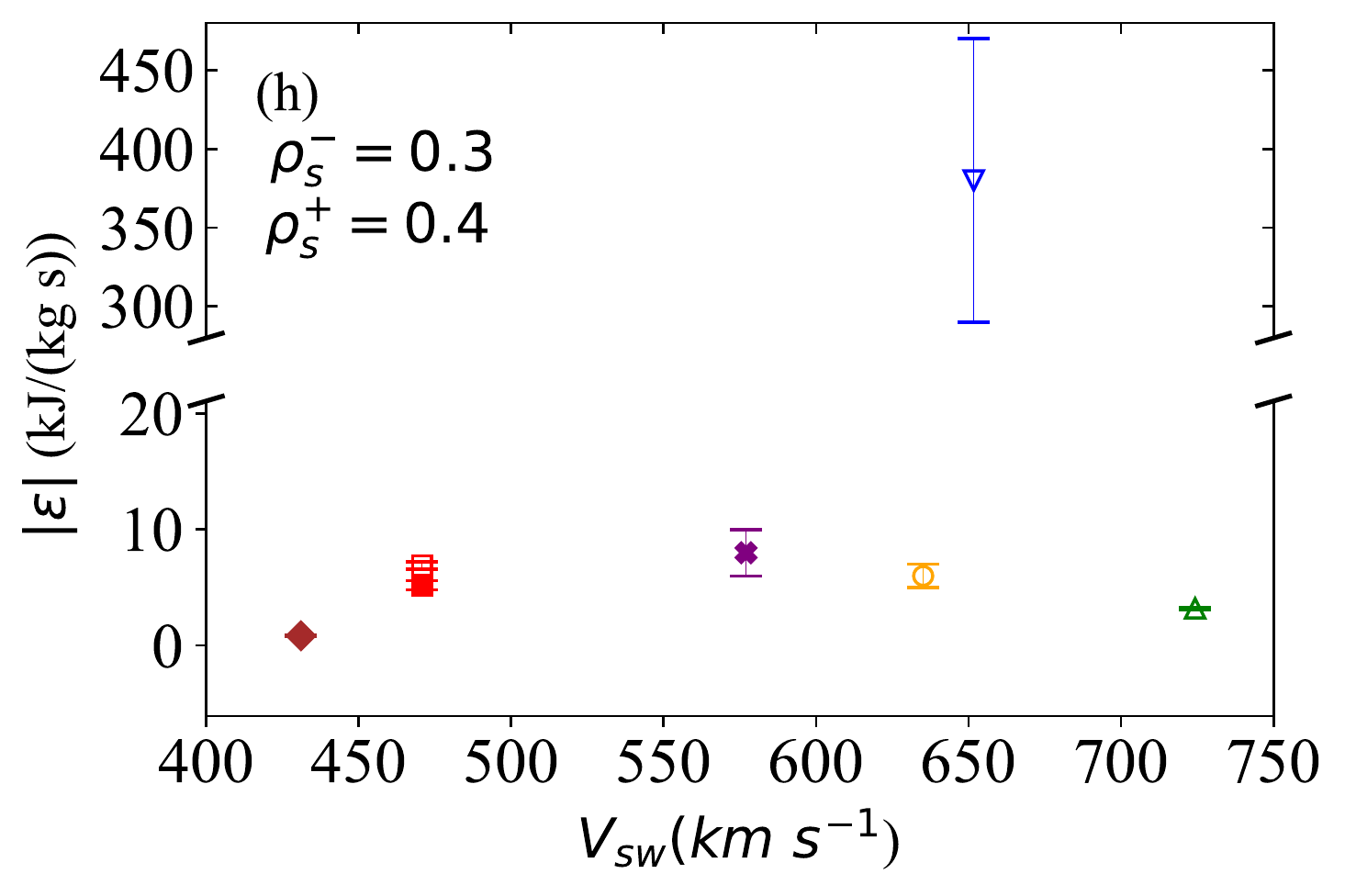}
            }

\caption{{\it Panels (a)-(c):} absolute values of the computed mean turbulent energy transfer rates, $|\varepsilon|$, versus spectral exponents $\alpha$ for $\BB$, $\bv$ and $\np$ in each region. 
Dotted vertical lines stand for the 5/3 and 3/2 for K41 \citep{Kolmogorov1941} and IK64 \citep{Iroshnikov1964,Kraichnan1965} spectra. 
{\it Panels (d)-(f):} $|\varepsilon|$ versus flatness exponents, $\kappa$, for $\BB$, $\bv$ and $\np$, in each region. 
{\it Panels (g)-(h):} $|\varepsilon|$ versus $|\sigmac|$ and $\vsw$, respectively. 
In all panels, error bars correspond to the linear fit uncertainties. 
The $y$-axis have been broken for better visualization, due to a much higher value of $|\varepsilon|$ in the sheath region (blue).
Legends in every panel (but g) display  two distinct Spearman coefficients, $\rho_{s}^{-}$ and $\rho_{s}^{+}$, corresponding to the two values (one negative and one positive, respectively) obtained for $\varepsilon$ for SW1 (panel (a) of Figure \ref{F-Yaglom panel}).}
   \label{F-epsilon panel}
   \end{figure}

\begin{figure}[h!]
\centering
\includegraphics[width=0.8\textwidth]{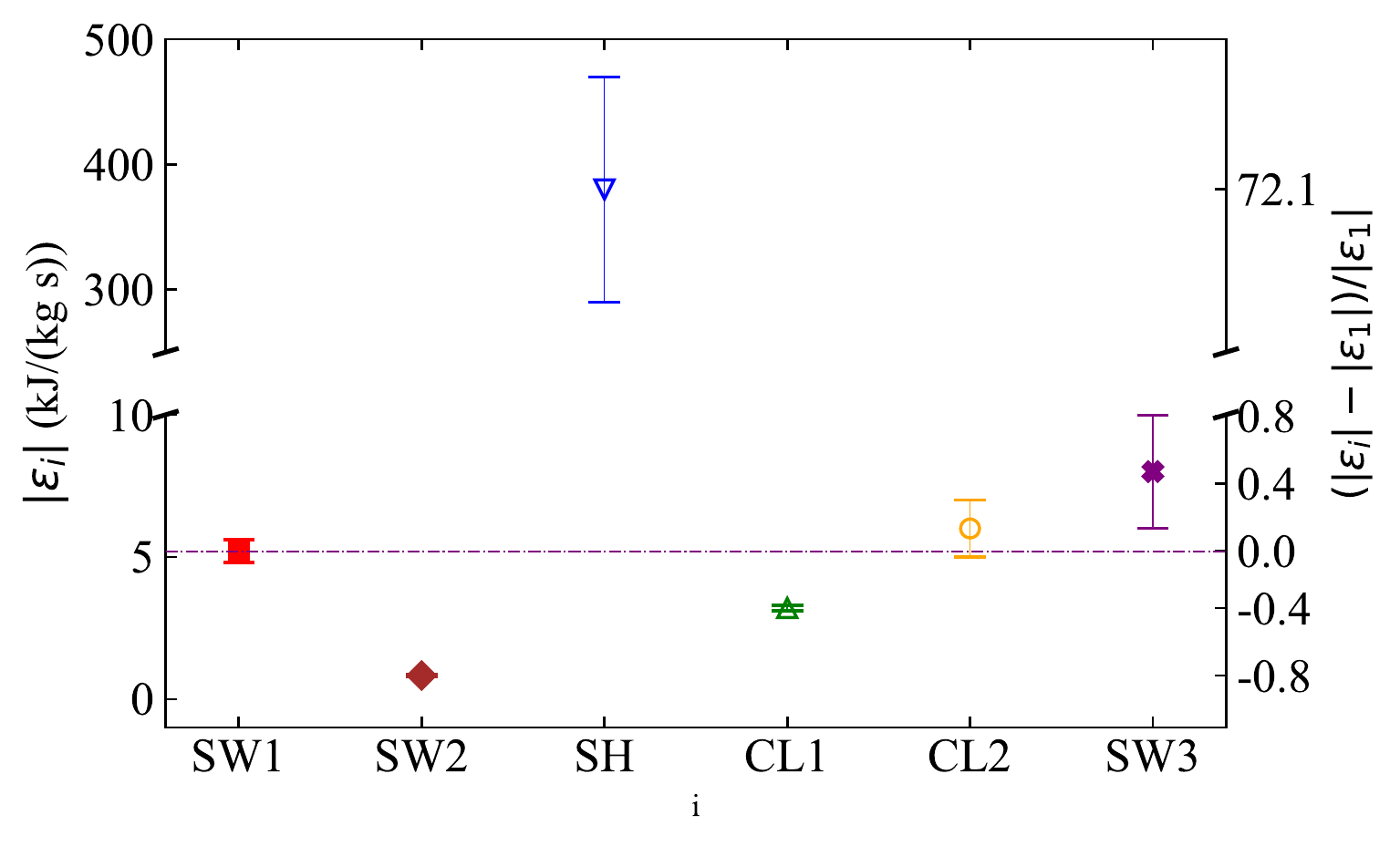}

\caption{Absolute value of the mean energy transfer rate, $|\epsilon_{i}|$, displayed for each sub-interval $i=\mathrm{\{SW1, SW2, SH, CL1, CL2, SW3\}}$ (colors and symbols as in previous figures). 
The right $y$-axis gives the normalized difference between each sub-interval and SW1, $(|\epsilon_{i}|-|\epsilon_{1}|)/|\epsilon_{1}|$, where $\epsilon_{1}$ stands for the positive value for SW1 (see panel a of Figure \ref{F-Yaglom panel}).
The horizontal line indicates $\epsilon_{1}$ (left y-axis) or 0\% variation with respect to $\epsilon_{1}$ (right y-axis). Notice that both $y$-axes have been split to easily display the higher value for the SH sub-interval. Error bars represent linear fit uncertainties.}
\label{F-normalized}
\end{figure}

Panels (a)-(h) in figure \ref{F-epsilon panel} show all computed energy transfer rates, plotted against the previously studied spectral indices (panels (a)-(c)), flatness scaling exponents (panels (d)-(f)) and $|\sigmac|$ and $\vsw$ (panels (g) and (h), respectively). 
The order of magnitude of $\varepsilon$ (units have been converted to kJ/kg s) is in agreement with values obtained in similar intervals by \citet{Sorriso-Valvo2021}. 
A notable exception is the sheath region (blue), in which the energy transfer rate is two orders of magnitude larger than in the other regions (note that, as a consequence, $y$-axis in Figure \ref{F-epsilon panel} have been split into two, so as to visually display the relative much higher value of the sheath region's $\varepsilon$). Such extreme value is likely due to the strong energy injection at the ICME shock. 
Furthermore, it is also related to the much higher intermittency for the velocity, as measured through the scaling exponent $\kappa$ of the flatness (see panel (e) in Figure \ref{F-F panel}). 
A similar result was previously observed in a sheath region, where the density (and not the velocity) displayed strong intermittency \citep{Sorriso-Valvo2021}. 
According to the results from those two case studies, the shocked ICME sheath regions are therefore rich in intermittent plasma structures (seen in velocity or in density) and, correspondingly, have a higher turbulent energy transfer. 
Such turbulent energy might be contributing to the plasma heating resulting from the strong compression in the sheath \citep{Yordanova2021}. 

Several Spearman correlation coefficients and their associated p-values were estimated for those pairs of parameters represented in Figure \ref{F-epsilon panel} (with the exception of panel (g)). The Spearman coefficients are displayed in each of those graphs, being denoted as $\rho_{s}^{-}$ or $\rho_{s}^{+}$, depending on which (negative or positive, respectively) value of $\varepsilon$ from the SW1 region has been taken into account (recall that SW1 provided two linear fits, corresponding to negative and positive values of Y in Equation \ref{Eq-PP law}; see panel (a) in Figure \ref{F-Yaglom panel}).
Particularly good correlations with the energy transfer rates were found for the magnetic field spectral exponent (panel (a)), the velocity spectral exponent (panel (b)) and for the velocity flatness scaling exponent (panel (e)). Moreover, the magnetic field flatness exponent also shows a good correlation if the outlier value of the sheath is excluded. 
This suggests that higher energy transfer rates are associated with steeper spectra and enhanced intermittency, both being indicative of a more developed turbulence. 

Finally, in Figure \ref{F-normalized} we explore the variations of the turbulence across the ICME structure. 
To this aim, the energy transfer rate, $\varepsilon_i$, is plotted versus a sub-interval index, $i=\mathrm{\{SW1, SW2, SH, CL1, CL2, SW3\}}$ (colors and symbols as in previous figures), representing the succession of regions studied in this paper. 
In the same figure, the right $y$-axis gives the normalized difference between each sub-interval and the reference value in the undisturbed solar wind preceding the ICME (SW1), $(|\epsilon_{i}|-|\epsilon_{1}|)/|\epsilon_{1}|$, where $\epsilon_{1}$ stands for the positive value for SW1 (see panel (a) of Figure \ref{F-Yaglom panel}).
The dashed horizontal line indicates the reference value, $\epsilon_{1}$ (left y-axis), or 0\% variation with respect to $\epsilon_{1}$ (right y-axis). 
The quiet SW2 and CL1 intervals have reduced cascade rate, associated with the extremely smooth profiles and low level of fluctuations in those regions. The high value in the sheath region has been already discussed, and shows clearly the increase of turbulence associated to the severe plasma compression and driven by the ICME shock.
On the other hand, the trailing portion of the cloud (CL2) and the solar wind in the wake of the ICME (SW3) show enhanced turbulent energy transfer rate, up to 50\% of the reference value. Although we cannot exclude that such enhancement stems from statistical fluctuations of the solar wind conditions, it is also possible that the ICME has injected additional energy in the large-scale fluctuations, which then feed a stronger turbulent cascade. 
A statistical study based on several events will be necessary in order to clarify the possibility of observing and measuring the modifications of solar wind turbulence due to ICME crossing.

\section{Conclusions}
     \label{S-Conclusion}

Several parameters associated to turbulent energy cascades have been studied in the case of the 12-14 September 2014 ICME and its preceding and trailing solar wind, measured by NASA's Wind spacecraft.
A structure function-based analysis, involving statistical scale-depending information of the fluctuations, was performed over a series of homogeneous sub-intervals corresponding to different sections of the ICME event. 
Computed spectral indices and flatness scaling exponents showed that turbulence was well-established within all regions. 
The Politano-Pouquet law was validated over several intervals, providing values of the mean energy transfer rates, which carry information on the turbulent energy flows across scales within the aforementioned regions. 
Experimental results showed that a linear scaling was found within and around the ICME cloud, even though isotropy and incompressibility were assumed while formulating the PP law. 
The mean energy transfer rate was exceptionally high in the sheath region, as a result of a powerful energy injection due to the arrival of the ICME’s shock. This is possibly related to the detected anomalously high value of the intermittency for the velocity, suggesting a prominent role of velocity structures (strong gradients and vortical flows) in determining the properties of the cascade. Turbulent energy transfer rates in other samples are magnetic field-dominated, as steeper spectra and larger intermittency indicate. 
Furthermore, high Spearman correlation coefficients were found while studying possible dependencies of the energy transfer rates on the spectral indices for the magnetic field and velocity, and on the flatness scaling exponent for the velocity, respectively. 
Most of the above observations are in accordance with those of the recent first study of the turbulent cascade within a ICME \citep{Sorriso-Valvo2021}. 
Finally, a preliminary observation of enhanced energy transfer rate in the trailing region and behind the ICME suggest a possible role of those structures in injecting further turbulence in the solar wind.

All theses results may help to improve our understanding of the turbulent properties of ICMEs and their expansion through interplanetary space, thus providing us with better tools to model their usually harmful interaction with the Earth's magnetosphere.

\begin{acks}
R.M.R. acknowledges financial support from the European Erasmus+ Programme for his stay at Uppsala University. 
He also thanks L.S.-V. and E.Y. for their useful advice and guidance during the realisation of this project, and IRF for having allowed him to carry out a research project with its members. 
L.S.-V. and E.Y. acknowledge funding from the SNSA grants 86/20 and 145/18. 
\end{acks}

\newpage
\bibliographystyle{spr-mp-sola}
\bibliography{CME}  

\begin{thebibliography}{46}
\ifx\bisbn     \undefined \def\bisbn  #1{ISBN #1}\fi
\ifx\binits    \undefined \def\binits#1{#1}\fi
\ifx\bauthor   \undefined \def\bauthor#1{#1}\fi
\ifx\batitle   \undefined \def\batitle#1{#1}\fi
\ifx\bjtitle   \undefined \def\bjtitle#1{\textit{#1}}\fi
\ifx\bvolume   \undefined \def\bvolume#1{\textbf{#1}}\fi
\ifx\byear     \undefined \def\byear#1{#1}\fi
\ifx\bissue    \undefined \def\bissue#1{#1}\fi
\ifx\bfpage    \undefined \def\bfpage#1{#1}\fi
\ifx\blpage    \undefined \def\blpage #1{#1}\fi
\ifx\burl      \undefined \def\burl#1{\textsf{#1}}\fi
\ifx\href      \undefined \def\href#1#2{\textsf{#2}}\fi
\ifx\betal     \undefined \def\betal{\textit{et al.}}\fi
\ifx\bctitle   \undefined \def\bctitle#1{#1}\fi
\ifx\beditor   \undefined \def\beditor#1{#1}\fi
\ifx\bbtitle   \undefined \def\bbtitle#1{\textit{#1}}\fi
\ifx\bedition  \undefined \def\bedition#1{#1}\fi
\ifx\bseriesno \undefined \def\bseriesno#1{\textbf{#1}}\fi
\ifx\blocation \undefined \def\blocation#1{#1}\fi
\ifx\bsertitle \undefined \def\bsertitle#1{\textit{#1}}\fi
\ifx\bsnm      \undefined \def\bsnm#1{#1}\fi
\ifx\bsuffix   \undefined \def\bsuffix#1{#1}\fi
\ifx\bparticle \undefined \def\bparticle#1{#1}\fi
\ifx\barticle  \undefined \def\barticle#1{}\fi
\ifx\binstitute  \undefined \def\binstitute#1{#1}\fi
\ifx\bpublisher  \undefined \def\bpublisher#1{#1}\fi
\ifx\doiurl    \undefined
  \def\doiurl#1{\href{http://dx.doi.org/#1}{\textsf{DOI}}}\fi
\ifx\arxivurl  \undefined
  \def\arxivurl#1{\href{http://arxiv.org/abs/#1}{\textsf{arXiv}}}\fi
\ifx\adsurl    \undefined
  \def\adsurl#1{\href{http://adsabs.harvard.edu/abs/#1}{\textsf{ADS}}}\fi
\ifx\botherref \undefined \def\botherref#1{}\fi
\ifx\url       \undefined \def\url#1{\textsf{#1}}\fi
\ifx\bchapter  \undefined \def\bchapter#1{}\fi
\ifx\bbook     \undefined \def\bbook#1{}\fi
\ifx\bcomment  \undefined \def\bcomment#1{#1}\fi
\ifx\oauthor   \undefined \def\oauthor#1{#1}\fi
\ifx\citeauthoryear \undefined\def \citeauthoryear#1{#1}\fi
\ifx\endbibitem\undefined \def\endbibitem{}\fi
\ifx\bconflocation  \undefined \def\bconflocation#1{#1} \fi

\bibitem[\protect\citeauthoryear{{Bothmer} and {Zhukov}}{2007}]{Bothmer2007}
\begin{bchapter}
\bauthor{\bsnm{{Bothmer}}, \binits{V.}},
\bauthor{\bsnm{{Zhukov}}, \binits{A.}}:
\byear{2007},
\bctitle{{The Sun as the prime source of space weather}}.
In: \beditor{\bsnm{{Bothmer}}, \binits{V.}},
\beditor{\bsnm{{Daglis}}, \binits{I.A.}} (eds.)
\bbtitle{Space Weather- Physics and Effects},
\bfpage{31}.
\doiurl{10.1007/978-3-540-34578-7_3}.
\adsurl{https://ui.adsabs.harvard.edu/abs/2007swpe.book...31B}.
\end{bchapter}
\endbibitem

\bibitem[\protect\citeauthoryear{{Bruno} and
  {Carbone}}{2013}]{BrunoCarbone2013}
\begin{barticle}
\bauthor{\bsnm{{Bruno}}, \binits{R.}},
\bauthor{\bsnm{{Carbone}}, \binits{V.}}:
\byear{2013},
\batitle{{The Solar Wind as a Turbulence Laboratory}}.
\bjtitle{{\it Liv. Rev. in Solar Phys.}}
\bvolume{10},
\bfpage{2}.
\end{barticle}
\endbibitem

\bibitem[\protect\citeauthoryear{Carbone and Sorriso-Valvo}{2014}]{Carbone2014}
\begin{barticle}
\bauthor{\bsnm{Carbone}, \binits{F.}},
\bauthor{\bsnm{Sorriso-Valvo}, \binits{L.}}:
\byear{2014},
\batitle{{Experimental analysis of intermittency in electrohydrodynamic
  instability}}.
\bjtitle{Eur. Phys. J. E}
\bvolume{37},
\bfpage{61}.
\end{barticle}
\endbibitem

\bibitem[\protect\citeauthoryear{Coburn \textit{et~al.}}{2015}]{Coburn2015}
\begin{barticle}
\bauthor{\bsnm{Coburn}, \binits{J.T.}},
\bauthor{\bsnm{Forman}, \binits{M.A.}},
\bauthor{\bsnm{Smith}, \binits{C.W.}},
\bauthor{\bsnm{Vasquez}, \binits{B.J.}},
\bauthor{\bsnm{Stawarz}, \binits{J.E.}}:
\byear{2015},
\batitle{Third-moment descriptions of the interplanetary turbulent cascade,
  intermittency and back transfer}.
\bjtitle{Philosophical Transactions of the Royal Society A: Mathematical,
  Physical and Engineering Sciences}
\bvolume{373}(\bissue{2041}),
\bfpage{20140150}.
\doiurl{10.1098/rsta.2014.0150}.
\burl{https://royalsocietypublishing.org/doi/abs/10.1098/rsta.2014.0150}.
\end{barticle}
\endbibitem

\bibitem[\protect\citeauthoryear{{Dal Lago}, Schwenn, and
  Gonzalez}{2003}]{Dallago2003}
\begin{barticle}
\bauthor{\bsnm{{Dal Lago}}, \binits{A.}},
\bauthor{\bsnm{Schwenn}, \binits{R.}},
\bauthor{\bsnm{Gonzalez}, \binits{W.D.}}:
\byear{2003},
\batitle{Relation between the radial speed and theexpansion speed of coronal
  mass ejections}.
\bjtitle{Advances in Space Research}
\bvolume{32}(\bissue{12}),
\bfpage{2637}.
\doiurl{https://doi.org/10.1016/j.asr.2003.03.012}.
\burl{https://www.sciencedirect.com/science/article/pii/S0273117703800765}.
\end{barticle}
\endbibitem

\bibitem[\protect\citeauthoryear{{Dudok de Wit}
  \textit{et~al.}}{2013}]{Dudok2013}
\begin{barticle}
\bauthor{\bsnm{{Dudok de Wit}}, \binits{T.}},
\bauthor{\bsnm{{Alexandrova}}, \binits{O.}},
\bauthor{\bsnm{{Furno}}, \binits{I.}},
\bauthor{\bsnm{{Sorriso-Valvo}}, \binits{L.}},
\bauthor{\bsnm{{Zimbardo}}, \binits{G.}}:
\byear{2013},
\batitle{{Methods for Characterising Microphysical Processes in Plasmas}}.
\bjtitle{\ssr}
\bvolume{178}(\bissue{2-4}),
\bfpage{665}.
\doiurl{10.1007/s11214-013-9974-9}.
\end{barticle}
\endbibitem

\bibitem[\protect\citeauthoryear{{Echer}, {Tsurutani}, and
  {Gonzalez}}{2013}]{Echer2013}
\begin{barticle}
\bauthor{\bsnm{{Echer}}, \binits{E.}},
\bauthor{\bsnm{{Tsurutani}}, \binits{B.T.}},
\bauthor{\bsnm{{Gonzalez}}, \binits{W.D.}}:
\byear{2013},
\batitle{{Interplanetary origins of moderate (-100 nT < Dst {\ensuremath{\leq}}
  -50 nT) geomagnetic storms during solar cycle 23 (1996-2008)}}.
\bjtitle{Journal of Geophysical Research (Space Physics)}
\bvolume{118}(\bissue{1}),
\bfpage{385}.
\doiurl{10.1029/2012JA018086}.
\adsurl{https://ui.adsabs.harvard.edu/abs/2013JGRA..118..385E}.
\end{barticle}
\endbibitem

\bibitem[\protect\citeauthoryear{Elsasser}{1950}]{Elsasser1950}
\begin{barticle}
\bauthor{\bsnm{Elsasser}, \binits{W.M.}}:
\byear{1950},
\batitle{The hydromagnetic equations}.
\bjtitle{Phys. Rev.}
\bvolume{79},
\bfpage{183}.
\doiurl{10.1103/PhysRev.79.183}.
\burl{https://link.aps.org/doi/10.1103/PhysRev.79.183}.
\end{barticle}
\endbibitem

\bibitem[\protect\citeauthoryear{Frisch}{1995}]{Frisch1995}
\begin{bbook}
\bauthor{\bsnm{Frisch}, \binits{U.}}:
\byear{1995},
\bbtitle{Turbulence: The legacy of a.n. kolmogorov},
\bpublisher{Cambridge University Press}, \blocation{???}.
\end{bbook}
\endbibitem

\bibitem[\protect\citeauthoryear{{Frisch}, {Sulem}, and
  {Nelkin}}{1978}]{Frisch1978}
\begin{barticle}
\bauthor{\bsnm{{Frisch}}, \binits{U.}},
\bauthor{\bsnm{{Sulem}}, \binits{P.-L.}},
\bauthor{\bsnm{{Nelkin}}, \binits{M.}}:
\byear{1978},
\batitle{{A simple dynamical model of intermittent fully developed
  turbulence}}.
\bjtitle{\jfm}
\bvolume{87},
\bfpage{719}.
\end{barticle}
\endbibitem

\bibitem[\protect\citeauthoryear{{Gui} \textit{et~al.}}{2011}]{Gui2011}
\begin{barticle}
\bauthor{\bsnm{{Gui}}, \binits{B.}},
\bauthor{\bsnm{{Shen}}, \binits{C.}},
\bauthor{\bsnm{{Wang}}, \binits{Y.}},
\bauthor{\bsnm{{Ye}}, \binits{P.}},
\bauthor{\bsnm{{Liu}}, \binits{J.}},
\bauthor{\bsnm{{Wang}}, \binits{S.}},
\bauthor{\bsnm{{Zhao}}, \binits{X.}}:
\byear{2011},
\batitle{{Quantitative Analysis of CME Deflections in the Corona}}.
\bjtitle{\solphys}
\bvolume{271}(\bissue{1-2}),
\bfpage{111}.
\doiurl{10.1007/s11207-011-9791-9}.
\adsurl{https://ui.adsabs.harvard.edu/abs/2011SoPh..271..111G}.
\end{barticle}
\endbibitem

\bibitem[\protect\citeauthoryear{{Heinemann}
  \textit{et~al.}}{2019}]{Heinemann2019}
\begin{barticle}
\bauthor{\bsnm{{Heinemann}}, \binits{S.G.}},
\bauthor{\bsnm{{Temmer}}, \binits{M.}},
\bauthor{\bsnm{{Farrugia}}, \binits{C.J.}},
\bauthor{\bsnm{{Dissauer}}, \binits{K.}},
\bauthor{\bsnm{{Kay}}, \binits{C.}},
\bauthor{\bsnm{{Wiegelmann}}, \binits{T.}},
\bauthor{\bsnm{{Dumbovi{\'c}}}, \binits{M.}},
\bauthor{\bsnm{{Veronig}}, \binits{A.M.}},
\bauthor{\bsnm{{Podladchikova}}, \binits{T.}},
\bauthor{\bsnm{{Hofmeister}}, \binits{S.J.}},
\bauthor{\bsnm{{Lugaz}}, \binits{N.}},
\bauthor{\bsnm{{Carcaboso}}, \binits{F.}}:
\byear{2019},
\batitle{{CME-HSS Interaction and Characteristics Tracked from Sun to Earth}}.
\bjtitle{\solphys}
\bvolume{294}(\bissue{9}),
\bfpage{121}.
\doiurl{10.1007/s11207-019-1515-6}.
\adsurl{https://ui.adsabs.harvard.edu/abs/2019SoPh..294..121H}.
\end{barticle}
\endbibitem

\bibitem[\protect\citeauthoryear{Hern{\'{a}}ndez
  \textit{et~al.}}{2021}]{Hernandez2021}
\begin{barticle}
\bauthor{\bsnm{Hern{\'{a}}ndez}, \binits{C.S.}},
\bauthor{\bsnm{Sorriso-Valvo}, \binits{L.}},
\bauthor{\bsnm{Bandyopadhyay}, \binits{R.}},
\bauthor{\bsnm{Chasapis}, \binits{A.}},
\bauthor{\bsnm{V{\'{a}}sconez}, \binits{C.L.}},
\bauthor{\bsnm{Marino}, \binits{R.}},
\bauthor{\bsnm{Pezzi}, \binits{O.}}:
\byear{2021},
\batitle{Impact of switchbacks on turbulent cascade and energy transfer rate in
  the inner heliosphere}.
\bjtitle{The Astrophysical Journal Letters}
\bvolume{922}(\bissue{1}),
\bfpage{L11}.
\doiurl{10.3847/2041-8213/ac36d1}.
\burl{https://doi.org/10.3847/2041-8213/ac36d1}.
\end{barticle}
\endbibitem

\bibitem[\protect\citeauthoryear{{Howard}}{2011}]{Howard2011}
\begin{bbook}
\bauthor{\bsnm{{Howard}}, \binits{T.}}:
\byear{2011},
\bbtitle{{Coronal Mass Ejections: An Introduction}}
\bseriesno{376}.
\doiurl{10.1007/978-1-4419-8789-1}.
\end{bbook}
\endbibitem

\bibitem[\protect\citeauthoryear{{Iroshnikov}}{1964}]{Iroshnikov1964}
\begin{barticle}
\bauthor{\bsnm{{Iroshnikov}}, \binits{P.S.}}:
\byear{1964},
\batitle{{Turbulence of a Conducting Fluid in a Strong Magnetic Field}}.
\bjtitle{\sovast}
\bvolume{7},
\bfpage{566}.
\end{barticle}
\endbibitem

\bibitem[\protect\citeauthoryear{{Kilpua}, {Koskinen}, and
  {Pulkkinen}}{2017}]{Kilpua2017}
\begin{barticle}
\bauthor{\bsnm{{Kilpua}}, \binits{E.}},
\bauthor{\bsnm{{Koskinen}}, \binits{H.E.J.}},
\bauthor{\bsnm{{Pulkkinen}}, \binits{T.I.}}:
\byear{2017},
\batitle{{Coronal mass ejections and their sheath regions in interplanetary
  space}}.
\bjtitle{Living Reviews in Solar Physics}
\bvolume{14}(\bissue{1}),
\bfpage{5}.
\doiurl{10.1007/s41116-017-0009-6}.
\end{barticle}
\endbibitem

\bibitem[\protect\citeauthoryear{{Kilpua} \textit{et~al.}}{2021}]{Kilpua2021}
\begin{barticle}
\bauthor{\bsnm{{Kilpua}}, \binits{E.K.J.}},
\bauthor{\bsnm{{Good}}, \binits{S.W.}},
\bauthor{\bsnm{{Ala-Lahti}}, \binits{M.}},
\bauthor{\bsnm{{Osmane}}, \binits{A.}},
\bauthor{\bsnm{{Fontaine}}, \binits{D.}},
\bauthor{\bsnm{{Hadid}}, \binits{L.}},
\bauthor{\bsnm{{Janvier}}, \binits{M.}},
\bauthor{\bsnm{{Yordanova}}, \binits{E.}}:
\byear{2021},
\batitle{{Statistical analysis of magnetic field fluctuations in CME-driven
  sheath regions}}.
\bjtitle{Frontiers in Astronomy and Space Sciences}
\bvolume{7},
\bfpage{109}.
\doiurl{10.3389/fspas.2020.610278}.
\end{barticle}
\endbibitem

\bibitem[\protect\citeauthoryear{{Kolmogorov}}{1941}]{Kolmogorov1941}
\begin{barticle}
\bauthor{\bsnm{{Kolmogorov}}, \binits{A.N.}}:
\byear{1941},
\batitle{{Dissipation of Energy in Locally Isotropic Turbulence}}.
\bjtitle{{\it Dokl. Akad. Nauk SSSR}}
\bvolume{32},
\bfpage{16}.
\end{barticle}
\endbibitem

\bibitem[\protect\citeauthoryear{{Kolmogorov}}{1962}]{Kolmogorov1962}
\begin{barticle}
\bauthor{\bsnm{{Kolmogorov}}, \binits{A.N.}}:
\byear{1962},
\batitle{{A refinement of previous hypotheses concerning the local structure of
  turbulence in a viscous incompressible fluid at high Reynolds number}}.
\bjtitle{{\jfm}}
\bvolume{13},
\bfpage{82}.
\end{barticle}
\endbibitem

\bibitem[\protect\citeauthoryear{{Kraichnan}}{1965}]{Kraichnan1965}
\begin{barticle}
\bauthor{\bsnm{{Kraichnan}}, \binits{R.H.}}:
\byear{1965},
\batitle{{Inertial-Range Spectrum of Hydromagnetic Turbulence}}.
\bjtitle{{\it Phys. of Fluids}}
\bvolume{8},
\bfpage{1385}.
\end{barticle}
\endbibitem

\bibitem[\protect\citeauthoryear{Lepping \textit{et~al.}}{1995}]{Lepping1995}
\begin{barticle}
\bauthor{\bsnm{Lepping}, \binits{R.P.}},
\bauthor{\bsnm{Acu\~na}, \binits{M.H.}},
\bauthor{\bsnm{Burlaga}, \binits{L.F.}},
\bauthor{\bparticle{et} \bsnm{al.}}:
\byear{1995},
\batitle{{The WIND magnetic field investigation}}.
\bjtitle{\ssr}
\bvolume{71}(\bissue{207}).
\doiurl{https://doi.org/10.1007/BF00751330}.
\end{barticle}
\endbibitem

\bibitem[\protect\citeauthoryear{{Lugaz} \textit{et~al.}}{2017}]{Lugaz2017}
\begin{barticle}
\bauthor{\bsnm{{Lugaz}}, \binits{N.}},
\bauthor{\bsnm{{Temmer}}, \binits{M.}},
\bauthor{\bsnm{{Wang}}, \binits{Y.}},
\bauthor{\bsnm{{Farrugia}}, \binits{C.J.}}:
\byear{2017},
\batitle{{The Interaction of Successive Coronal Mass Ejections: A Review}}.
\bjtitle{\solphys}
\bvolume{292}(\bissue{4}),
\bfpage{64}.
\doiurl{10.1007/s11207-017-1091-6}.
\adsurl{https://ui.adsabs.harvard.edu/abs/2017SoPh..292...64L}.
\end{barticle}
\endbibitem

\bibitem[\protect\citeauthoryear{Marino and Sorriso-Valvo}{2023}]{Marino2023}
\begin{barticle}
\bauthor{\bsnm{Marino}, \binits{R.}},
\bauthor{\bsnm{Sorriso-Valvo}, \binits{L.}}:
\byear{2023},
\batitle{Scaling laws for the energy transfer in space plasma turbulence}.
\bjtitle{Phys. Rep.}
\bvolume{X},
\bfpage{X}.
\end{barticle}
\endbibitem

\bibitem[\protect\citeauthoryear{{Marino} \textit{et~al.}}{2012}]{Marino2012}
\begin{barticle}
\bauthor{\bsnm{{Marino}}, \binits{R.}},
\bauthor{\bsnm{{Sorriso-Valvo}}, \binits{L.}},
\bauthor{\bsnm{{D'Amicis}}, \binits{R.}},
\bauthor{\bsnm{{Carbone}}, \binits{V.}},
\bauthor{\bsnm{{Bruno}}, \binits{R.}},
\bauthor{\bsnm{{Veltri}}, \binits{P.}}:
\byear{2012},
\batitle{{On the Occurrence of the Third-order Scaling in High Latitude Solar
  Wind}}.
\bjtitle{\apj}
\bvolume{750}(\bissue{1}),
\bfpage{41}.
\doiurl{10.1088/0004-637X/750/1/41}.
\end{barticle}
\endbibitem

\bibitem[\protect\citeauthoryear{Marino \textit{et~al.}}{2022}]{Marino2022}
\begin{barticle}
\bauthor{\bsnm{Marino}, \binits{R.}},
\bauthor{\bsnm{Feraco}, \binits{F.}},
\bauthor{\bsnm{Primavera}, \binits{L.}},
\bauthor{\bsnm{Pumir}, \binits{A.}},
\bauthor{\bsnm{Pouquet}, \binits{A.}},
\bauthor{\bsnm{Rosenberg}, \binits{D.}},
\bauthor{\bsnm{Mininni}, \binits{P.D.}}:
\byear{2022},
\batitle{Turbulence generation by large-scale extreme vertical drafts and the
  modulation of local energy dissipation in stably stratified geophysical
  flows}.
\bjtitle{Phys. Rev. Fluids}
\bvolume{7},
\bfpage{033801}.
\doiurl{10.1103/PhysRevFluids.7.033801}.
\burl{https://link.aps.org/doi/10.1103/PhysRevFluids.7.033801}.
\end{barticle}
\endbibitem

\bibitem[\protect\citeauthoryear{{Matthaeus} and {Velli}}{2011}]{Matthaeus2011}
\begin{barticle}
\bauthor{\bsnm{{Matthaeus}}, \binits{W.H.}},
\bauthor{\bsnm{{Velli}}, \binits{M.}}:
\byear{2011},
\batitle{{Who Needs Turbulence?. A Review of Turbulence Effects in the
  Heliosphere and on the Fundamental Process of Reconnection}}.
\bjtitle{\ssr}
\bvolume{160},
\bfpage{145}.
\end{barticle}
\endbibitem

\bibitem[\protect\citeauthoryear{Oughton and Matthaeus}{2020}]{Oughton2020}
\begin{barticle}
\bauthor{\bsnm{Oughton}, \binits{S.}},
\bauthor{\bsnm{Matthaeus}, \binits{W.H.}}:
\byear{2020},
\batitle{Critical balance and the physics of magnetohydrodynamic turbulence}.
\bjtitle{The Astrophysical Journal}
\bvolume{897}(\bissue{1}),
\bfpage{37}.
\doiurl{10.3847/1538-4357/ab8f2a}.
\burl{https://doi.org/10.3847/1538-4357/ab8f2a}.
\end{barticle}
\endbibitem

\bibitem[\protect\citeauthoryear{{Parker}}{1958}]{Parker1958}
\begin{barticle}
\bauthor{\bsnm{{Parker}}, \binits{E.N.}}:
\byear{1958},
\batitle{{Dynamics of the Interplanetary Gas and Magnetic Fields.}}
\bjtitle{\apj}
\bvolume{128},
\bfpage{664}.
\end{barticle}
\endbibitem

\bibitem[\protect\citeauthoryear{Politano and Pouquet}{1998}]{Politano1998}
\begin{barticle}
\bauthor{\bsnm{Politano}, \binits{H.}},
\bauthor{\bsnm{Pouquet}, \binits{A.}}:
\byear{1998},
\batitle{{von Karman-Howarth equation for magnetohydrodynamics and its
  consequences on third-order longitudinal structure and correlation
  functions}}.
\bjtitle{Phys. Rev. E}
\bvolume{57}(\bissue{1}),
\bfpage{R21}.
\doiurl{10.1103/PhysRevE.57.R21}.
\end{barticle}
\endbibitem

\bibitem[\protect\citeauthoryear{Pulkkinen
  \textit{et~al.}}{2007}]{Pulkkinen2007}
\begin{barticle}
\bauthor{\bsnm{Pulkkinen}, \binits{T.I.}},
\bauthor{\bsnm{Palmroth}, \binits{M.}},
\bauthor{\bsnm{Tanskanen}, \binits{E.I.}},
\bauthor{\bsnm{Ganushkina}, \binits{N.Y.}},
\bauthor{\bsnm{Shukhtina}, \binits{M.A.}},
\bauthor{\bsnm{Dmitrieva}, \binits{N.P.}}:
\byear{2007},
\batitle{Solar wind—magnetosphere coupling: A review of recent results}.
\bjtitle{Journal of Atmospheric and Solar-Terrestrial Physics}
\bvolume{69}(\bissue{3}),
\bfpage{256}.
\bcomment{Global Aspects of Magnetosphere-Ionosphere Coupling}.
\doiurl{https://doi.org/10.1016/j.jastp.2006.05.029}.
\burl{https://www.sciencedirect.com/science/article/pii/S1364682606002653}.
\end{barticle}
\endbibitem

\bibitem[\protect\citeauthoryear{{Quijia} \textit{et~al.}}{2021}]{Quijia2021}
\begin{barticle}
\bauthor{\bsnm{{Quijia}}, \binits{P.}},
\bauthor{\bsnm{{Fraternale}}, \binits{F.}},
\bauthor{\bsnm{{Stawarz}}, \binits{J.E.}},
\bauthor{\bsnm{{V{\'a}sconez}}, \binits{C.L.}},
\bauthor{\bsnm{{Perri}}, \binits{S.}},
\bauthor{\bsnm{{Marino}}, \binits{R.}},
\bauthor{\bsnm{{Yordanova}}, \binits{E.}},
\bauthor{\bsnm{{Sorriso-Valvo}}, \binits{L.}}:
\byear{2021},
\batitle{{Comparing turbulence in a Kelvin-Helmholtz instability region across
  the terrestrial magnetopause}}.
\bjtitle{\mnras}
\bvolume{503}(\bissue{4}),
\bfpage{4815}.
\doiurl{10.1093/mnras/stab319}.
\end{barticle}
\endbibitem

\bibitem[\protect\citeauthoryear{{Schwenn} \textit{et~al.}}{2005}]{Schwenn2005}
\begin{barticle}
\bauthor{\bsnm{{Schwenn}}, \binits{R.}},
\bauthor{\bsnm{{dal Lago}}, \binits{A.}},
\bauthor{\bsnm{{Huttunen}}, \binits{E.}},
\bauthor{\bsnm{{Gonzalez}}, \binits{W.D.}}:
\byear{2005},
\batitle{{The association of coronal mass ejections with their effects near the
  Earth}}.
\bjtitle{Annales Geophysicae}
\bvolume{23}(\bissue{3}),
\bfpage{1033}.
\doiurl{10.5194/angeo-23-1033-2005}.
\adsurl{https://ui.adsabs.harvard.edu/abs/2005AnGeo..23.1033S}.
\end{barticle}
\endbibitem

\bibitem[\protect\citeauthoryear{{Smith} \textit{et~al.}}{2009}]{Smith2009}
\begin{barticle}
\bauthor{\bsnm{{Smith}}, \binits{C.W.}},
\bauthor{\bsnm{{Stawarz}}, \binits{J.E.}},
\bauthor{\bsnm{{Vasquez}}, \binits{B.J.}},
\bauthor{\bsnm{{Forman}}, \binits{M.A.}},
\bauthor{\bsnm{{MacBride}}, \binits{B.T.}}:
\byear{2009},
\batitle{{Turbulent Cascade at 1 AU in High Cross-Helicity Flows}}.
\bjtitle{\prl}
\bvolume{103}(\bissue{20}),
\bfpage{201101}.
\doiurl{10.1103/PhysRevLett.103.201101}.
\end{barticle}
\endbibitem

\bibitem[\protect\citeauthoryear{{Sorriso-Valvo}
  \textit{et~al.}}{1999}]{Sorriso-Valvo1999}
\begin{barticle}
\bauthor{\bsnm{{Sorriso-Valvo}}, \binits{L.}},
\bauthor{\bsnm{{Carbone}}, \binits{V.}},
\bauthor{\bsnm{{Veltri}}, \binits{P.}},
\bauthor{\bsnm{{Consolini}}, \binits{G.}},
\bauthor{\bsnm{{Bruno}}, \binits{R.}}:
\byear{1999},
\batitle{{Intermittency in the solar wind turbulence through probability
  distribution functions of fluctuations}}.
\bjtitle{\grl}
\bvolume{26},
\bfpage{1801}.
\end{barticle}
\endbibitem

\bibitem[\protect\citeauthoryear{{Sorriso-Valvo}
  \textit{et~al.}}{2007}]{Sorriso-Valvo2007}
\begin{barticle}
\bauthor{\bsnm{{Sorriso-Valvo}}, \binits{L.}},
\bauthor{\bsnm{{Marino}}, \binits{R.}},
\bauthor{\bsnm{{Carbone}}, \binits{V.}},
\bauthor{\bsnm{{Noullez}}, \binits{A.}},
\bauthor{\bsnm{{Lepreti}}, \binits{F.}},
\bauthor{\bsnm{{Veltri}}, \binits{P.}},
\bauthor{\bsnm{{Bruno}}, \binits{R.}},
\bauthor{\bsnm{{Bavassano}}, \binits{B.}},
\bauthor{\bsnm{{Pietropaolo}}, \binits{E.}}:
\byear{2007},
\batitle{{Observation of Inertial Energy Cascade in Interplanetary Space
  Plasma}}.
\bjtitle{\prl}
\bvolume{99}(\bissue{11}),
\bfpage{115001}.
\end{barticle}
\endbibitem

\bibitem[\protect\citeauthoryear{{Sorriso-Valvo}
  \textit{et~al.}}{2018}]{Sorriso-Valvo2018}
\begin{barticle}
\bauthor{\bsnm{{Sorriso-Valvo}}, \binits{L.}},
\bauthor{\bsnm{{Carbone}}, \binits{F.}},
\bauthor{\bsnm{{Perri}}, \binits{S.}},
\bauthor{\bsnm{{Greco}}, \binits{A.}},
\bauthor{\bsnm{{Marino}}, \binits{R.}},
\bauthor{\bsnm{{Bruno}}, \binits{R.}}:
\byear{2018},
\batitle{{On the Statistical Properties of Turbulent Energy Transfer Rate in
  the Inner Heliosphere}}.
\bjtitle{\solphys}
\bvolume{293}(\bissue{1}),
\bfpage{10}.
\doiurl{10.1007/s11207-017-1229-6}.
\end{barticle}
\endbibitem

\bibitem[\protect\citeauthoryear{{Sorriso-Valvo}
  \textit{et~al.}}{2021}]{Sorriso-Valvo2021}
\begin{barticle}
\bauthor{\bsnm{{Sorriso-Valvo}}, \binits{L.}},
\bauthor{\bsnm{{Yordanova}}, \binits{E.}},
\bauthor{\bsnm{{Dimmock}}, \binits{A.P.}},
\bauthor{\bsnm{{Telloni}}, \binits{D.}}:
\byear{2021},
\batitle{{Turbulent cascade and energy transfer rate in a solar coronal mass
  ejection}}.
\bjtitle{\apjl}
\bvolume{919},
\bfpage{L30}.
\doiurl{10.3847/2041-8213/ac26c5}.
\end{barticle}
\endbibitem

\bibitem[\protect\citeauthoryear{Stawarz \textit{et~al.}}{2011}]{Stawarz2011}
\begin{barticle}
\bauthor{\bsnm{Stawarz}, \binits{J.E.}},
\bauthor{\bsnm{Vasquez}, \binits{B.J.}},
\bauthor{\bsnm{Smith}, \binits{C.W.}},
\bauthor{\bsnm{Forman}, \binits{M.A.}},
\bauthor{\bsnm{Klewicki}, \binits{J.}}:
\byear{2011},
\batitle{{THIRD} {MOMENTS} {AND} {THE} {ROLE} {OF} {ANISOTROPY} {FROM}
  {VELOCITY} {SHEAR} {IN} {THE} {SOLAR} {WIND}}.
\bjtitle{The Astrophysical Journal}
\bvolume{736}(\bissue{1}),
\bfpage{44}.
\doiurl{10.1088/0004-637x/736/1/44}.
\burl{https://doi.org/10.1088/0004-637x/736/1/44}.
\end{barticle}
\endbibitem

\bibitem[\protect\citeauthoryear{{Taylor}}{1938}]{Taylor1938}
\begin{barticle}
\bauthor{\bsnm{{Taylor}}, \binits{G.I.}}:
\byear{1938},
\batitle{{The Spectrum of Turbulence}}.
\bjtitle{{\it Royal Society of London Proceedings Series A}}
\bvolume{164},
\bfpage{476}.
\end{barticle}
\endbibitem

\bibitem[\protect\citeauthoryear{{Temmer}}{2021}]{Temmer2021}
\begin{barticle}
\bauthor{\bsnm{{Temmer}}, \binits{M.}}:
\byear{2021},
\batitle{{Space weather: the solar perspective}}.
\bjtitle{Living Reviews in Solar Physics}
\bvolume{18}(\bissue{1}),
\bfpage{4}.
\doiurl{10.1007/s41116-021-00030-3}.
\adsurl{https://ui.adsabs.harvard.edu/abs/2021LRSP...18....4T}.
\end{barticle}
\endbibitem

\bibitem[\protect\citeauthoryear{Verdini \textit{et~al.}}{2015}]{Verdini2015b}
\begin{barticle}
\bauthor{\bsnm{Verdini}, \binits{A.}},
\bauthor{\bsnm{Grappin}, \binits{R.}},
\bauthor{\bsnm{Hellinger}, \binits{P.}},
\bauthor{\bsnm{Landi}, \binits{S.}},
\bauthor{\bsnm{Müller}, \binits{W.C.}}:
\byear{2015},
\batitle{{ANISOTROPY} {OF} {THIRD}-{ORDER} {STRUCTURE} {FUNCTIONS} {IN} {MHD}
  {TURBULENCE}}.
\bjtitle{The Astrophysical Journal}
\bvolume{804}(\bissue{2}),
\bfpage{119}.
\doiurl{10.1088/0004-637x/804/2/119}.
\burl{https://doi.org/10.1088/0004-637x/804/2/119}.
\end{barticle}
\endbibitem

\bibitem[\protect\citeauthoryear{Viall and Borovsky}{2020}]{Viall2020}
\begin{barticle}
\bauthor{\bsnm{Viall}, \binits{N.M.}},
\bauthor{\bsnm{Borovsky}, \binits{J.E.}}:
\byear{2020},
\batitle{Nine outstanding questions of solar wind physics}.
\bjtitle{Journal of Geophysical Research: Space Physics}
\bvolume{125}(\bissue{7}),
\bfpage{e2018JA026005}.
\doiurl{https://doi.org/10.1029/2018JA026005}.
\end{barticle}
\endbibitem

\bibitem[\protect\citeauthoryear{{Wang}, {Hoeksema}, and
  {Liu}}{2020}]{Wang2020}
\begin{barticle}
\bauthor{\bsnm{{Wang}}, \binits{J.}},
\bauthor{\bsnm{{Hoeksema}}, \binits{J.T.}},
\bauthor{\bsnm{{Liu}}, \binits{S.}}:
\byear{2020},
\batitle{{The Deflection of Coronal Mass Ejections by the Ambient Coronal
  Magnetic Field Configuration}}.
\bjtitle{Journal of Geophysical Research (Space Physics)}
\bvolume{125}(\bissue{8}),
\bfpage{e27530}.
\doiurl{10.1029/2019JA027530}.
\adsurl{https://ui.adsabs.harvard.edu/abs/2020JGRA..12527530W}.
\end{barticle}
\endbibitem

\bibitem[\protect\citeauthoryear{{Wang} \textit{et~al.}}{2004}]{Wang2004}
\begin{barticle}
\bauthor{\bsnm{{Wang}}, \binits{Y.}},
\bauthor{\bsnm{{Shen}}, \binits{C.}},
\bauthor{\bsnm{{Wang}}, \binits{S.}},
\bauthor{\bsnm{{Ye}}, \binits{P.}}:
\byear{2004},
\batitle{{Deflection of coronal mass ejection in the interplanetary medium}}.
\bjtitle{\solphys}
\bvolume{222}(\bissue{2}),
\bfpage{329}.
\doiurl{10.1023/B:SOLA.0000043576.21942.aa}.
\adsurl{https://ui.adsabs.harvard.edu/abs/2004SoPh..222..329W}.
\end{barticle}
\endbibitem

\bibitem[\protect\citeauthoryear{Yordanova
  \textit{et~al.}}{2021}]{Yordanova2021}
\begin{barticle}
\bauthor{\bsnm{Yordanova}, \binits{E.}},
\bauthor{\bsnm{Vörös}, \binits{Z.}},
\bauthor{\bsnm{Sorriso-Valvo}, \binits{L.}},
\bauthor{\bsnm{Dimmock}, \binits{A.P.}},
\bauthor{\bsnm{Kilpua}, \binits{E.}}:
\byear{2021},
\batitle{A possible link between turbulence and plasma heating}.
\bjtitle{The Astrophysical Journal}
\bvolume{921}(\bissue{1}),
\bfpage{65}.
\doiurl{10.3847/1538-4357/ac1942}.
\burl{https://doi.org/10.3847/1538-4357/ac1942}.
\end{barticle}
\endbibitem

\bibitem[\protect\citeauthoryear{Zurbuchen and
  Richardson}{2006}]{Zurbuchen2006}
\begin{bbook}
\bauthor{\bsnm{Zurbuchen}, \binits{T.H.}},
\bauthor{\bsnm{Richardson}, \binits{I.G.}}:
\byear{2006},
\bbtitle{In-Situ Solar Wind and Magnetic Field Signatures of Interplanetary
  Coronal Mass Ejections},
\bpublisher{Springer},
\blocation{New York, NY},
\bfpage{31}.
\bisbn{978-0-387-45088-9}.
\doiurl{10.1007/978-0-387-45088-9$_$3}.
\end{bbook}
\endbibitem

\end{thebibliography}

\end{article} 

\end{document}